\documentclass[11pt]{article}
\usepackage{color}
\usepackage{subfigure}
\usepackage{amsmath}
\usepackage{amssymb}
\usepackage{graphicx,color}
\usepackage{cite}
\usepackage{enumerate}
\usepackage{amsthm}% theorems and proofs
\usepackage{amsfonts,mathrsfs}
\usepackage{geometry}
\usepackage{ifthen}
\usepackage{graphicx}
\usepackage{epstopdf}
\usepackage{float}
\usepackage{bm}
\usepackage[caption=false]{subfig}
\usepackage[utf8]{inputenc}
\usepackage{etoolbox}

\begin{document}

\title{\bf Geometric-resource bounds and permutation-averaged success probabilities for the WZP algorithm}

\vskip0.1in
\author{\small Linlin Ye$^1$, Linmao Wang$^1$, Zhaoqi Wu$^1$\thanks{Corresponding author. E-mail: wuzhaoqi\_conquer@163.com},
Shao-Ming Fei$^{2,3}$\\
{\small\it  1. Department of Mathematics, Nanchang University,
Nanchang 330031, P R China}\\
{\small\it  2. School of Mathematical Sciences, Capital Normal University, Beijing 100048, P R China}\\
{\small\it  3. Max-Planck-Institute for Mathematics in the Sciences,
04103 Leipzig, Germany} }

\date{}
\maketitle

\noindent {\bf Abstract} {\small }\\
Solving linear systems of equations is a fundamental task in machine learning and optimization. In this work, we extend the original WZP algorithm from real symmetric linear systems to complex Hermitian linear systems. We then study the success probability of the modified WZP algorithm from a resource-theoretic perspective. By expressing the postselection success probability as the expectation value of a positive operator, we derive upper and lower bounds of the maximal success probability over separable, incoherent, and real pure states. These bounds further establish connections between the maximal success probability and geometric entanglement, coherence, and imaginarity. We further show that, under permutation averaging, the averaged success probability is an affine function of the coherence fraction, with its dependence determined by the alignment between the success operator and the equal-superposition state. We also investigate the preprocessed WZP algorithm under subgroup-constrained preprocessing. For the two-level family, the optimal success probability is determined by the maximal projection weight onto the orbit of the target projector. Our results provide a unified perspective on the WZP success probability under different operational constraints.

\vskip0.2in

\noindent {\bf 1. Introduction}\\\hspace*{\fill}\\
Quantum coherence was initially introduced to explain light interference. It arises from the quantum superposition principle and serves as a crucial resource in quantum information processing \cite{SEC,GRJ,SMO}. Its rigorous quantification has been systematically established through the resource-theoretic framework~\cite{TB,YXD}, which has inspired a variety of coherence measures, including the robustness of coherence~\cite{RobustnessCoh} and geometric coherence~\cite{SAS}. These frameworks have further stimulated extensive research on coherence dynamics and related phenomena~\cite{QXG,DSB,SDZ,TRB,JWZ,YWF-Grover,YWF-Linear,XDY,YPYJ,AMV,CRM,YWF-Simon,YWF-Shor,
WZQH,WZZL,YWZ-Circuit,WZZF}, as well as on their connections with other quantum resources~\cite{ECM,JMB,ZXY,YYX}. In addition, significant progress has been made in extending coherence quantification to infinite-dimensional systems~\cite{JXP,YRZ}, thereby broadening the applicability of coherence theory within quantum information science.
Recent works have investigated the geometric coherence for several classes of mixed states and characterized their nearest incoherent states \cite{Zhao2024GeoCoherence}.

Quantum entanglement is a distinctive nonclassical feature of composite quantum systems \cite{Horodecki2009,Bruss2002}. It is widely regarded as a fundamental resource in quantum information processing, and plays an important role in many quantum tasks, including tests of nonlocality \cite{Bell1964}, quantum communication \cite{Bennett1992} and cryptography \cite{Ekert1991}.
Geometric entanglement quantifies entanglement through the maximal overlap between a target state and separable states \cite{Shimony1995}, which provides an intuitive description and is particularly useful for multipartite systems \cite{Wei2003}. It is also closely related to other entanglement measures and can be estimated experimentally by entanglement witnesses \cite{Guhne2007}. Geometric entanglement has also been used to study distance–entanglement tradeoffs in quantum error-correcting codes, showing the dependence on the code family and the entanglement measure \cite{Bravyi2025QECGeometric}.

The indispensable role of complex numbers in quantum mechanics is well recognized, as their influence permeates the foundations and development of the theory \cite{Namiki1991,Ricardo2020,David2018}. To explore the significance of complex numbers within quantum systems and the potential advantages they confer, Hickey and Gour~\cite{Hickey2018} proposed the resource theory of imaginarity.
The geometric imaginarity quantifies how far a quantum state is from the set of real states \cite{Wu2021}.
Closely related to this line, geometric-like imaginarity and related imaginarity quantifiers have also been introduced and analyzed from the viewpoints of quantification and state conversion \cite{Guo2025GeoLikeImaginarity,WuWu2025CTP,GMG-AQT,GMG-PRA,MLi2026,Guo2026SetImaginarity}. The imaginarity of quantum operations, together with the preservation and generation of imaginarity in distributed scenarios, has also been investigated recently \cite{WW-Operations,WMH-PRA}.

Solving systems of linear equations is a fundamental computational task that underlies many problems in science and engineering. The HHL algorithm~\cite{HAW} can be used to solve linear systems of the form $A|x\rangle=|b\rangle$ for given matrix $A$ and vector $|b\rangle$, but it requires the matrix $A$ to be sparse and well-conditioned. These requirements, however, are often violated in the large-scale and ill-conditioned linear systems that arise in modern machine-learning applications, and limit the practical applicability of the HHL algorithm. Motivated by this limitation, Kerenidis et al.~\cite{Kerenidis} proposed a quantum singular-value estimation algorithm (QSVE) based on a tree-structured quantum random-access memory (QRAM). Building on their QRAM-based framework, Wossnig et al.~\cite{Wossnig} subsequently developed a quantum linear-systems solver capable of handling dense matrices, commonly referred to as the WZP algorithm. Recent surveys continue to discuss WZP as a representative QSVE-based approach for dense matrices \cite{Papagiannis2025QSolvers}. Beyond linear-system solving, singular-value-based and, more broadly, spectral primitives have continued to play important roles in quantum algorithm design, as reflected in recent advances in optimal-depth quantum state preparation \cite{ZhangLiYuan2022QSP}, quantum algorithms for fidelity susceptibility \cite{ZhangYuan2026FS}, and sum-of-squares spectral amplification for quantum simulation \cite{King2026SOSSA}. Related recent advances also include non-Hermitian eigenproblems and nonlinear property estimation \cite{Zhang2025NonHermitian,Zhang2026NonlinearProperties}.

Meanwhile, quantum algorithms for linear systems have continued to develop along several directions. Nguyen et al. \cite{Nguyen2022DenseKernels} proposed block-encoding methods for dense and full-rank kernel matrices by using hierarchical matrices, extending quantum numerical linear algebra beyond sparse settings. Bravo-Prieto et al. \cite{BravoPrieto2023VQLS} introduced the variational quantum linear solver for near-term devices, while Ghisoni et al. \cite{Ghisoni2026SQLS} proposed a resource-efficient quantum linear solver. Related advances in quantum machine learning and optimization also highlight the broader utility of quantum linear-algebraic and spectral primitives \cite{Gong2024QPSO,Gong2024QKNNDC,Gong2024QCNNVQC,Huang2023HQCNN}.

Against this background, we first extend the original WZP algorithm
from real symmetric linear systems to complex Hermitian linear systems
and introduce the modified WZP algorithm. We then study its success
probability under restricted input classes and permutation-averaged
input ensembles. For restricted input classes, the maximal success
probability is related to geometric overlaps with the dominant
spectral subspaces, while under permutation averaging, the averaged
success probability is an affine function of the coherence fraction.
We further investigate the preprocessed WZP algorithm under
subgroup-constrained preprocessing, where the optimal success
probability for the two-level family is determined by the maximal
projection weight onto the orbit of the target projector.

The remainder of this paper is organized as follows. In Section 2, we review the WZP algorithm and recall the geometric resource quantifiers. In Section 3, we introduce the modified WZP algorithm for complex Hermitian linear systems and derive bounds on the maximal success probability over separable, incoherent and real pure states. In Section 4, we study group-averaged success probabilities and derive an exact permutation-averaged formula in terms of the coherence fraction. In Section 5, we introduce the preprocessed WZP algorithm and study its optimal success probability under subgroup-constrained preprocessing. Finally, Section 6 concludes the paper.

\vskip0.2in
\noindent {\bf 2. WZP algorithm and geometric measures of resources}\\

We first briefly review the WZP algorithm and related geometric measures of resources.
Let $A\in\mathbb{R}^{N\times N}$ be a Hermitian matrix with spectral decomposition
$A=\sum_{i=1}^{N}\lambda_i|v_i\rangle\langle v_i|$,
where the eigenvalues satisfy
$\lambda_i\in[-1,-1/\kappa]\cup[1/\kappa,1]$ with $\kappa$ being the condition number of $A$.
Given a normalized vector $|b\rangle=\sum_{i=1}^{N}\beta_i|v_i\rangle$,
the goal is to solve the linear system $A|x\rangle=|b\rangle$.
Unlike the HHL algorithm, the WZP algorithm does not require $A$ to be sparse.
Since $A$ is Hermitian, its singular values are $\sigma_i=|\lambda_i|$.

The WZP algorithm is based on quantum singular value estimation (QSVE) \cite{Kerenidis,Wossnig}.
Let $X=\sum_{j=1}^{r}\sigma_j |\mu_j\rangle\langle \nu_j|$ be the singular value decomposition of $X\in\mathbb{R}^{m\times n}$,
and let
$|y\rangle=\sum_{j=1}^{r}\alpha_j |\nu_j\rangle$.
Given an absolute precision parameter $\delta>0$, QSVE performs the transformation
\begin{equation}\label{eq:qsve_transform}
\sum_{j=1}^{r}\alpha_j |\nu_j\rangle |0\rangle
\longrightarrow
\sum_{j=1}^{r}\alpha_j |\nu_j\rangle |\bar{\sigma}_j\rangle,
\end{equation}
where each estimate satisfies $|\bar{\sigma}_j-\sigma_j|\leq \delta$ with high probability, and the runtime is $O\!\left(\|X\|_F\,\operatorname{polylog}(mn)/\delta\right)$.
The main steps of the WZP algorithm are summarized below \cite{Wossnig}.

\noindent (i) Choose a precision parameter $\delta<1/(2\kappa)$ and set $\mu=1/\kappa$.
Apply the singular value estimation to $A$ and $A+\mu I$,
$$
|\psi_1\rangle
=\sum_i \beta_i |v_i\rangle_A ||\lambda_i|\rangle_B ||\lambda_i+\mu|\rangle_C.
$$

\noindent (ii) Introduce an ancilla register $D$ and define
$f_i=1$ if $|\lambda_i+\mu|\le |\lambda_i|$, and $f_i=0$ otherwise.
After a controlled phase operation, the state becomes
$$
|\psi_2\rangle
=
\sum_i (-1)^{f_i}\beta_i |v_i\rangle_A ||\lambda_i|\rangle_B ||\lambda_i+\mu|\rangle_C |f_i\rangle_D.
$$
\noindent (iii) Add another ancilla register $E$ and perform a controlled rotation with parameter $\gamma$
satisfying $0<\gamma\leq \min_j |\lambda_j(A)|$.
After uncomputing the registers $B$, $C$ and $D$, one obtains
$$
|\psi_3\rangle
=
\sum_i (-1)^{f_i}\beta_i |v_i\rangle_A
\left(
\frac{\gamma}{|\lambda_i|}|0\rangle
+
\sqrt{1-\frac{\gamma^2}{|\lambda_i|^2}}\,|1\rangle
\right)_E.
$$
\noindent (iv) Conditioned on measuring $E$ in the state $|0\rangle$, the post-measurement state is
$$
|\psi_4\rangle
=
\frac{1}{\sqrt{P_0}}
\sum_i (-1)^{f_i}\beta_i \frac{\gamma}{|\lambda_i|}|v_i\rangle,
$$
where the success probability is
$
P_0
=
\sum_i |\beta_i|^2\frac{\gamma^2}{|\lambda_i|^2}.
$
Equivalently,
\begin{equation}\label{eq:M_operator}
P_0(|b\rangle)
=
\langle b|M|b\rangle,
\qquad
M:=\sum_{i=1}^{N}\frac{\gamma^2}{|\lambda_i|^2}|v_i\rangle\langle v_i|.
\end{equation}

Concerning the geometric measures of resource, we adopt the root-fidelity,
$F(\rho,\sigma)=\mathrm{Tr}\sqrt{\sqrt{\rho}\,\sigma\sqrt{\rho}}$,
for two quantum states $\rho$ and $\sigma$\cite{MA}. In particular, if $\rho=|\psi\rangle\langle\psi|$ is pure, then
$
F(|\psi\rangle\langle\psi|,\sigma)=\sqrt{\langle\psi|\sigma|\psi\rangle}.
$
Let $\mathcal{I}$ denote the set of incoherent states. The geometric coherence is defined by\cite{SAS}
\begin{equation}\label{eq:geom_coherence}
C_g(\rho)
=
1-\left[\max_{\delta\in\mathcal{I}}F(\rho,\delta)\right]^2.
\end{equation}
For a pure state $|\psi\rangle=\sum_{i}\alpha_i |i\rangle$, one has
\begin{equation}\label{eq:geom_coherence_pure}
C_g(|\psi\rangle)
=
1-\max_{|\phi\rangle\in\mathcal{I}_p}|\langle\phi|\psi\rangle|^2
=
1-\max_i |\alpha_i|^2,
\end{equation}
where $\mathcal{I}_p$ denotes the set of incoherent pure states.

For a pure $n$-qubit state $|\psi\rangle$, the geometric entanglement is defined by\cite{Shimony1995}
\begin{equation}\label{eq:geom_entanglement}
E_g(|\psi\rangle)
=
1-\max_{|\phi\rangle\in\mathcal{S}_p}|\langle\phi|\psi\rangle|^2,
\end{equation}
where $\mathcal{S}_p$ denotes the set of fully separable pure states
$
|\phi\rangle=\bigotimes_{s=1}^{n}|\phi_s\rangle,
$
with each $|\phi_s\rangle$ being a single-qubit pure state.

For a pure state $|\psi\rangle$, the geometric imaginarity is defined by\cite{Wu2021}
\begin{equation}\label{eq:geom_imaginarity}
I_g(|\psi\rangle)
=
1-\max_{|\phi\rangle\in\mathcal{R}}|\langle\phi|\psi\rangle|^2,
\end{equation}
where $\mathcal{R}$ denotes the set of pure states with real coefficients.
\vskip0.2in

\noindent {\bf 3. Success probability of the modified WZP algorithm under restricted input classes}

We extend the operator formulation of the WZP success probability to the complex Hermitian setting and derive bounds on the maximal success probability when the admissible input states are restricted to separable, incoherent, or real pure states. To this end, we first introduce a modified WZP algorithm for complex Hermitian linear systems by specifying the complex-valued QRAM realization required by the QSVE procedure. We then formulate the restricted-input optimization problem in a unified geometric way.

\smallskip

\noindent {\bf 3.1 Modified WZP algorithm}

The discussion in Section 2 was specialized to the original WZP algorithm for linear systems
$A|x\rangle=|b\rangle, A\in\mathbb{R}^{N\times N},$
where $A$ is real symmetric. The WZP algorithm is based on quantum singular-value estimation (QSVE) and a tree-structured quantum random-access memory (QRAM) data structure. In contrast to the HHL algorithm, which requires the coefficient matrix to be sparse, the WZP algorithm uses matrix information stored in QRAM to efficiently prepare quantum states associated with the rows of the matrix, thereby allowing dense matrices to be treated \cite{Kerenidis,Wossnig}.

Complex-valued linear systems arise naturally in quantum linear algebra, since quantum states, quantum operators, and Hamiltonians are generally defined over complex Hilbert spaces. It is therefore relevant to consider linear systems with
$A\in\mathbb{C}^{N\times N},
A=A^\dagger.$
Although a Hermitian matrix still has real eigenvalues, its matrix elements may carry nontrivial complex phases. Extending the WZP algorithm from real symmetric to complex Hermitian matrices thus requires these phases to be properly incorporated into the QRAM data structure and the associated state preparation procedure.

Related techniques for handling complex matrix data in QRAM-based quantum algorithms have appeared in previous works. Wang and Wossnig \cite{Wang2020} considered QRAM-based state preparation for the simulation of non-sparse Hermitian Hamiltonians, where the phase information of complex matrix elements must be retained in the data structure. Li et al. \cite{LiWLF2019} employed QSVE in a quantum data fitting algorithm for non-sparse matrices and discussed the treatment of complex Hermitian matrices. These works provide useful ingredients for handling complex matrix data, but their specific tasks and constructions are different from the complex Hermitian WZP linear system setting considered here. Motivated by these techniques, and combining them with the original QSVE/WZP construction \cite{Kerenidis,Wossnig}, we make explicit below the QRAM data structure and the corresponding isometries required for complex Hermitian linear systems.

The modification concerns the QRAM encoding of complex entries and the preparation of conjugated row states. The construction below ensures that the QSVE factorization recovers $A/\|A\|_F$.

{\bf Data Structure}  Assume that $A\in\mathbb{C}^{N\times N}$ is Hermitian and invertible. Writing
$A_{ij}=r_{ij}e^{\mathrm{i}\phi_{ij}},
r_{ij}=|A_{ij}|,$
we adapt the binary tree QRAM data structure used in the real valued QSVE construction to complex entries. For each row $i$, we construct a binary tree whose $j$-th leaf stores the pair
$\left(|A_{ij}|^2,-\phi_{ij}\right),$
while each internal node stores the sum of the squared moduli associated with the leaves in its subtree. The root of the tree corresponding to row $i$ therefore contains
\begin{equation*}
\sum_{j=1}^N |A_{ij}|^2=\|A_i\|_2^2,
\end{equation*}
where $A_i$ denotes the $i$th row of $A$.
A second binary tree is constructed over the row index $i$, with its leaves storing $\|A_i\|_2^2$. Its root therefore contains
\begin{equation*}
\sum_{i=1}^N\|A_i\|_2^2=\|A\|_F^2.
\end{equation*}
Compared with the real-valued data structure, the essential additional ingredient is the phase information $-\phi_{ij}$ stored at the leaves. This negative phase allows the QRAM to prepare the conjugated row state
$|A_i^*\rangle
:=
\frac{1}{\|A_i\|_2}
\sum_{j=1}^N A_{ij}^*|j\rangle.$

Based on this data structure, the following unitaries and their inverses can be implemented
\begin{equation*}
U_{\mathcal{T}}:\ |i\rangle|0\rangle
\mapsto
|i\rangle|A_i^*\rangle,
\qquad
|A_i^*\rangle
:=
\frac{1}{\|A_i\|_2}
\sum_{j=1}^N A_{ij}^*|j\rangle,
\end{equation*}
and
\begin{equation*}
U_{\mathcal{S}}:\ |0\rangle|j\rangle
\mapsto
|A_F\rangle|j\rangle,
\qquad
|A_F\rangle
:=
\frac{1}{\|A\|_F}
\sum_{i=1}^N\|A_i\|_2|i\rangle.
\end{equation*}
Here $\|\cdot\|_2$ denotes the Euclidean norm of a row vector and $\|\cdot\|_F$ denotes the Frobenius norm.
We denote the corresponding QRAM isometries by
\begin{equation*}
\mathcal{T}|i\rangle
=
|i\rangle|A_i^*\rangle,
\qquad
\mathcal{S}|j\rangle
=
|A_F\rangle|j\rangle.
\end{equation*}
Since $\{|i\rangle\}_{i=1}^N$ is an orthonormal basis and each $|A_i^*\rangle$ is normalized, one has
$\mathcal{T}^\dagger\mathcal{T}=I_N$ and
$\mathcal{S}^\dagger\mathcal{S}=I_N.$
The crucial point is that the above conjugated row-state construction preserves the desired factorization of $A$. For any $i,j\in\{1,\ldots,N\}$,
\begin{align*}
\langle i|\mathcal{T}^\dagger \mathcal{S}|j\rangle
=
\langle i,A_i^*|A_F,j\rangle
=
\frac{1}{\|A_i\|_2\|A\|_F}
\sum_{k,\ell=1}^N
A_{ik}\,\|A_\ell\|_2\,
\langle i|\ell\rangle
\langle k|j\rangle
=
\frac{A_{ij}}{\|A\|_F}.
\end{align*}
Hence,
$\mathcal{T}^\dagger\mathcal{S}
=
\frac{A}{\|A\|_F}.$
The conjugation in $|A_i^*\rangle$ is essential for this relation. Indeed,
$
\langle A_i^*|
=
\frac{1}{\|A_i\|_2}
\sum_{j=1}^N A_{ij}\langle j|,
$
which produces the matrix element $A_{ij}$ in $\mathcal{T}^\dagger\mathcal{S}$. If instead the unconjugated row state
$
|A_i\rangle
=
\frac{1}{\|A_i\|_2}
\sum_{j=1}^N A_{ij}|j\rangle
$
were used, the same construction would yield
$
\mathcal{T}^\dagger\mathcal{S}
=
\frac{A^*}{\|A\|_F}
$
rather than $A/\|A\|_F$. Thus, the conjugated row-state preparation is necessary for the QSVE construction to encode the original matrix $A$.
With this modification, the WZP walk operator retains the same form as in the real case
\begin{equation*}
W
=
(2\mathcal{T}\mathcal{T}^\dagger-I)
(2\mathcal{S}\mathcal{S}^\dagger-I).
\end{equation*}
Since
$
\mathcal{T}^\dagger\mathcal{S}
=
\frac{A}{\|A\|_F},
$
the corresponding QSVE procedure extracts the singular value information of $A$. Moreover, because $A$ is Hermitian, it admits the spectral decomposition
$
A
=
\sum_{i=1}^N
\lambda_i|v_i\rangle\langle v_i|,
\lambda_i\in\mathbb{R},
$
and its singular values are therefore $|\lambda_i|$.
The sign recovery procedure of the original WZP algorithm, which compares the QSVE outputs associated with $A$ and $A+\mu I$, can consequently be carried over to the complex Hermitian case. The subsequent controlled rotation, uncomputation, and postselection steps remain unchanged.

\paragraph{Restricted-input success probability}
From now on we work with the idealized exact spectral success operator introduced in Eq.~(\ref{eq:M_operator}),
\begin{equation*}
M=\gamma^2A^{-2}=\sum_{\ell=1}^{L}\omega_\ell \Lambda_\ell,
\qquad
\omega_1>\omega_2>\cdots>\omega_L>0,
\end{equation*}
where $\omega_1,\dots,\omega_L$ are the distinct eigenvalues of $M$, and $\Lambda_\ell$ is the spectral projector onto the eigenspace associated with $\omega_\ell$. For any pure input state $|b\rangle$, the idealized success probability of the modified WZP algorithm is
\begin{equation}\label{eq:sec3_success_probability}
P_0(|b\rangle)=\langle b|M|b\rangle.
\end{equation}

We consider variants for which the admissible input pure states are restricted to a prescribed compact set $S$, while all subsequent steps of the protocol remain unchanged. Define
\begin{equation}\label{eq:sec3_Pmax_general}
P_{\max}^S
:=
\max_{|b\rangle\in S}\langle b|M|b\rangle.
\end{equation}
We are mainly interested in the following three input classes.

\smallskip

\noindent
(i) {\it Fully separable states}
Denote $N=2^n$ and let
$
\mathcal{S}_p
:=
\left\{
|b\rangle=|b_1\rangle\otimes\cdots\otimes|b_n\rangle
:
\langle b_j|b_j\rangle=1
\right\}
$
be the set of fully separable pure $n$-qubit states. Then the corresponding maximal success probability is
\begin{equation*}
P_{\max}^{\mathrm{sep}}
:=
\max_{|b\rangle\in\mathcal{S}_p} P_0(|b\rangle).
\end{equation*}
\smallskip

\noindent
(ii) {\it Incoherent pure states}
Let $\mathcal{I}_p:=\{|x\rangle\}_{x=1}^{N}$ be
the set of computational-basis pure states. Then
\begin{equation*}
P_{\max}^{\mathrm{inc}}
:=
\max_{|b\rangle\in\mathcal{I}_p} P_0(|b\rangle)
=
\max_{x} P_0(|x\rangle).
\end{equation*}
\smallskip

\noindent
(iii) {\it Real pure states.}
Let
$\mathcal{R}
:=
\left\{
e^{\mathrm{i}\theta}\sum_{x=1}^{N} r_x |x\rangle
:
\theta\in\mathbb{R},\ r_x\in\mathbb{R},\ \sum_{x=1}^{N} r_x^2=1
\right\}$ denote
the set of pure states that are real in the computational basis up to a global phase. The corresponding maximal success probability is
\begin{equation*}
P_{\max}^{\mathrm{real}}
:=
\max_{|b\rangle\in\mathcal{R}} P_0(|b\rangle).
\end{equation*}
These restricted variants quantify how much the optimal success probability of the modified WZP algorithm survives when the admissible input is forced to lie inside a prescribed set of free states.

\smallskip

\noindent {\bf 3.2 Bounding the maximal success probability in the modified WZP algorithm}

For $\ell=1,\dots,L$, define the cumulative spectral projector
$\Pi_\ell=\sum_{t=1}^{\ell}\Lambda_t.$
Thus $\Pi_\ell$ projects onto the direct sum of the eigenspaces corresponding to the $\ell$ largest distinct eigenvalue levels of $M$, with $\Pi_L=I$. Let $S$ be a compact admissible set of pure states. We define
$\Gamma_S(\ell)=\max_{|b\rangle\in S}\langle b|\Pi_\ell|b\rangle,$ $\ell=1,\dots,L.$ We have the following theorem, see proof in Appendix A.

\smallskip

\noindent {\bf Theorem 1 }
The maximal success probability satisfies
\begin{equation}\label{eq:sec3_hierarchy_main}
\max_{1\le \ell\le L}
\Bigl[
\omega_L+(\omega_\ell-\omega_L)\Gamma_S(\ell)
\Bigr]
\le
P_{\max}^S
\le
\omega_L+\sum_{\ell=1}^{L-1}(\omega_\ell-\omega_{\ell+1})\Gamma_S(\ell).
\end{equation}
The upper bound in Eq.~(\ref{eq:sec3_hierarchy_main}) is attained if and only if there exists an admissible state $|b_*\rangle\in S$ such that $\langle b_*|\Pi_\ell|b_*\rangle=\Gamma_S(\ell)$ for all $\ell=1,\dots,L-1$.
Moreover, let
$
\ell^* \in \arg\max_{1 \leq \ell \leq L}
\left[\omega_L + (\omega_\ell - \omega_L)\Gamma_S(\ell)\right].
$
The lower bound in Eq.~(\ref{eq:sec3_hierarchy_main}) is attained if and only if
$
D_{\ell^*}(b)
\leq
(\omega_{\ell^*} - \omega_L)
\left[\Gamma_S(\ell^*) - s_{\ell^*}(b)\right],
\forall |b\rangle \in S,
$
where
$
s_{\ell^*}(b)
=
\langle b|\Pi_{\ell^*}|b\rangle
$
and
$
D_{\ell^*}(b)
=
\sum_{t<\ell^*}(\omega_t - \omega_{\ell^*})q_t
+
\sum_{\ell^*<t<L}(\omega_t - \omega_L)q_t.$

For $\mathcal{F}=\mathcal{S}_p$, $\mathcal{F}=\mathcal{I}_p$ and $\mathcal{F}=\mathcal{R}$, the results correspond to the geometric entanglement $E_g$, the geometric coherence $C_g$ and the geometric imaginarity $I_g$, respectively. For any closed pure-state set $\mathcal{F}$, define
\begin{equation*}
\Delta_{\mathcal{F}}(|\psi\rangle)=1-\max_{|\phi\rangle\in\mathcal{F}} |\langle\phi|\psi\rangle|^2.
\end{equation*}

\smallskip

\noindent {\bf Corollary 1 }
Assume that the largest eigenvalue $\omega_1$ of $M$ is nondegenerate, and $|v_1\rangle$ is the corresponding normalized eigenvector. Then for any closed admissible pure-state set $\mathcal{F}$, it holds that
\begin{equation}\label{eq:sec3_corollary_general}
\omega_L+(\omega_1-\omega_L)\bigl(1-\Delta_{\mathcal{F}}(|v_1\rangle)\bigr)
\le
P_{\max}^{\mathcal{F}}
\le
\omega_2+(\omega_1-\omega_2)\bigl(1-\Delta_{\mathcal{F}}(|v_1\rangle)\bigr).
\end{equation}
In particular, we have
\begin{equation}\label{eq:sec3_cor_sep}
\omega_L+(\omega_1-\omega_L)\bigl(1-E_g(|v_1\rangle)\bigr)
\le
P_{\max}^{\mathrm{sep}}
\le
\omega_2+(\omega_1-\omega_2)\bigl(1-E_g(|v_1\rangle)\bigr),
\end{equation}
\begin{equation}\label{eq:sec3_cor_inc}
\omega_L+(\omega_1-\omega_L)\bigl(1-C_g(|v_1\rangle)\bigr)
\le
P_{\max}^{\mathrm{inc}}
\le
\omega_2+(\omega_1-\omega_2)\bigl(1-C_g(|v_1\rangle)\bigr),
\end{equation}
and
\begin{equation}\label{eq:sec3_cor_real}
\omega_L+(\omega_1-\omega_L)\bigl(1-I_g(|v_1\rangle)\bigr)
\le
P_{\max}^{\mathrm{real}}
\le
\omega_2+(\omega_1-\omega_2)\bigl(1-I_g(|v_1\rangle)\bigr).
\end{equation}
$\it{Proof}$. Since the leading eigenspace is one-dimensional, $\Pi_1=|v_1\rangle\langle v_1|$, we have
\[
\Gamma_{\mathcal{F}}(1)
=
\max_{|b\rangle\in\mathcal{F}} \langle b|\Pi_1|b\rangle
=
\max_{|b\rangle\in\mathcal{F}} |\langle b|v_1\rangle|^2
=
1-\Delta_{\mathcal{F}}(|v_1\rangle).
\]
The lower bound follows from Eq.~(9) by choosing $\ell=1$. For the upper bound, the spectral ordering gives
$$
M \leq \omega_2 I + (\omega_1-\omega_2)\Lambda_1.
$$
Taking the expectation value and maximizing over $\mathcal{F}$ proves Eq.~(10).
Eqs.~(11)--(13) follow immediately from the definitions of $E_g$, $C_g$ and $I_g$.\qed

The upper bound in Corollary 1 is attained if and only if there exists $|b_*\rangle \in \mathcal{F}$ such that
$
|\langle v_1|b_*\rangle|^2
=
1-\Delta_{\mathcal{F}}(|v_1\rangle),
|b_*\rangle \in \operatorname{Ran}(\Lambda_1+\Lambda_2).$
The lower bound in Corollary 1 is attained if and only if
$\sum_{t=2}^{L-1}(\omega_t-\omega_L)q_t(b)
\leq
(\omega_1-\omega_L)
\left[
1-\Delta_{\mathcal{F}}(|v_1\rangle)
-
|\langle v_1|b\rangle|^2
\right]
$
for every $|b\rangle \in \mathcal{F}$.

\smallskip

\noindent {\bf Remark 1}
(1) If $M$ has only two distinct eigenvalues, namely,
$M=\omega_2 I+(\omega_1-\omega_2)\Pi,$
where $\Pi$ is the spectral projector onto the top eigenspace of $M$, then
\begin{equation}\label{eq:sec3_two_level_exact}
P_{\max}^S
=
\omega_2+(\omega_1-\omega_2)\max_{|b\rangle\in S}\langle b|\Pi|b\rangle.
\end{equation}
Thus, for the two-level spectral case, the restricted-input problem for the modified WZP algorithm is solved completely by the maximal overlap with the top spectral subspace.\\
(2) If $A$ is real symmetric, then $M=\gamma^2 A^{-2}$ has a real unit eigenvector associated with its largest eigenvalue. Hence,
\begin{equation}\label{eq:sec3_real_special_case}
P_{\max}^{\mathrm{real}}=\omega_1.
\end{equation}
(3) Because $\mathcal{I}_p$ consists of computational basis states, one always has the identity,
\begin{equation}\label{eq:sec3_inc_exact}
P_{\max}^{\mathrm{inc}}
=
\max_{x}\langle x|M|x\rangle
=
\max_{x} M_{xx}.
\end{equation}

\smallskip

Theorem~1 applies to general spectra, including degenerate eigenspaces,
through overlaps with cumulative spectral projectors. Its upper bound
is attained precisely when a single admissible input maximizes all
the relevant overlaps simultaneously. When the leading eigenspace
is one-dimensional, Corollary~1 expresses its maximal overlap in terms
of the corresponding geometric resource measure of the leading
eigenvector. For two spectral levels, the bounds coincide and give
the exact optimum.

\vskip0.2in

\noindent {\bf 4. Permutation-averaged success probability of the modified WZP algorithm}

In this section, we fix the coefficient matrix and average the WZP success probability over group-generated right-hand sides, specializing to computational-basis permutations. The resulting average is affine in the coherence fraction, and its slope can change sign with the eigenvector alignment even at fixed spectrum and condition number.

\smallskip

\noindent {\bf 4.1 Permutation averaging and coherence fraction}

Permutation averaging is motivated by the operational role of permutations in quantum algorithms. A permutation of the computational basis represents a reversible rearrangement of basis encoded data without changing the set of amplitudes. For example, in Shor's order finding algorithm \cite{Shor1997}, modular multiplication acts as a permutation of computational basis states, whose cyclic structure encodes the period extracted by quantum phase estimation. Thus, permutations provide a natural framework for studying how quantum algorithms respond to reversible rearrangements of input data.

For the WZP algorithm, the success probability depends not only on the spectrum of $A$, but also on the geometric alignment of the right hand side $|b\rangle$ with the spectral directions of $A$. Therefore, considering only a particular $|b\rangle$ makes it difficult to distinguish which properties arise from the specific coordinate arrangement of the input and which reflect more stable structural features. The permutation group $S_N$ provides a natural symmetrization: it only rearranges the components of $|b\rangle$ in the computational basis without changing the set of these components. Averaging over the entire permutation orbit removes the dependence on a particular labeling of the computational basis, thereby allowing us to identify which structural properties of the input state remain relevant to the performance of the WZP algorithm under all such rearrangements.

From now on, let $A\in\mathbb{C}^{N\times N}$ be a nonsingular Hermitian matrix
with $N\geq 2$.
To make the dependence on $A$ explicit, we write the success operator as
\begin{equation}\label{eq:sec4_MA}
M_A:=\gamma^2A^{-2},
\end{equation}
which satisfies $0<M_A\leq I$. For any normalized input state $|b\rangle$,
the WZP postselection success probability is
\begin{equation*}
P_0(|b\rangle;A)=\langle b|M_A|b\rangle.
\end{equation*}
Throughout this section, $A$ remains fixed while the input is varied.
Different right-hand sides generally correspond to distinct linear-system
instances rather than equivalent representations of a single fixed equation.

Let $G$ be a finite or compact group with a continuous unitary representation $g\mapsto U_g$ on $\mathbb{C}^N$. For a normalized input state $|b\rangle$, define its group orbit by\cite{Bartlett2007}
\begin{equation}\label{eq:sec4_orbit}
\mathcal{O}_G(b):=\{U_g|b\rangle:g\in G\}.
\end{equation}
The average success probability over this orbit is
\begin{equation}\label{eq:sec4_group_average}
\overline{P}_G(|b\rangle;A)
:=\int_G P_0(U_g|b\rangle;A)\,dg,
\end{equation}
where $dg$ denotes normalized Haar measure. For a finite group, the integral is understood as $|G|^{-1}\sum_{g\in G}$. Define the corresponding $G$-twirling map by\cite{Bartlett2007}
\begin{equation}\label{eq:sec4_twirling}
\mathcal{T}_G(X):=\int_G U_g^\dagger XU_g\,dg.
\end{equation}
It follows directly that
\begin{equation*}
\overline{P}_G(|b\rangle;A)
=\langle b|\mathcal{T}_G(M_A)|b\rangle.
\end{equation*}

We now derive the structure of the twirled operator using the irreducible decomposition of the group representation. The proof is given in Appendix B.

\smallskip

\noindent {\bf Lemma 1}
Consider the decomposition of the representation into inequivalent irreducible sectors,
$\mathbb{C}^N\simeq\bigoplus_\alpha\left(\mathbb{C}^{m_\alpha}\otimes\mathcal{H}_\alpha\right),$
$U_g=\bigoplus_\alpha\left(I_{m_\alpha}\otimes U_g^{(\alpha)}\right)$,
where $d_\alpha:=\dim\mathcal{H}_\alpha$, $m_\alpha$ is the multiplicity, and $\Pi_\alpha$ is the projector onto the corresponding sector. For any operator $X$, one has
\begin{equation}\label{eq:sec4_general_twirling}
\mathcal{T}_G(X)
=\bigoplus_\alpha
\left[
\frac{1}{d_\alpha}\operatorname{Tr}_{\mathcal{H}_\alpha}
\left(\Pi_\alpha X\Pi_\alpha\right)\otimes I_{\mathcal{H}_\alpha}
\right].
\end{equation}
Then the averaged success probability satisfies
\begin{equation}\label{eq:sec4_general_success}
\overline P_G(|b\rangle;A)
=\sum_\alpha\operatorname{Tr}\left[K_\alpha(A)\rho_\alpha^{(b)}\right].
\end{equation}
where $K_\alpha(A):=\frac{1}{d_\alpha}\operatorname{Tr}_{\mathcal{H}_\alpha}
\left(\Pi_\alpha M_A\Pi_\alpha\right)$ and
$\rho_\alpha^{(b)}:=\operatorname{Tr}_{\mathcal{H}_\alpha}
\left(\Pi_\alpha|b\rangle\langle b|\Pi_\alpha\right).$

When the representation is multiplicity-free, define
\begin{equation*}
\mu_\alpha(A):=\frac{\operatorname{Tr}(\Pi_\alpha M_A)}{d_\alpha},
\qquad
p_\alpha(|b\rangle):=\langle b|\Pi_\alpha|b\rangle.
\end{equation*}
Then
\begin{equation}\label{eq:sec4_multiplicity_free}
\overline{P}_G(|b\rangle;A)
=\sum_\alpha\mu_\alpha(A)p_\alpha(|b\rangle).
\end{equation}

We now take $G=S_N$ and let it act on the computational basis through the natural permutation representation,
\begin{equation*}
P_\pi|x\rangle=|\pi(x)\rangle,
\qquad \pi\in S_N.
\end{equation*}
Define the equal-superposition state
\begin{equation}\label{eq:sec4_eta}
|\eta\rangle:=\frac{1}{\sqrt{N}}\sum_{x=1}^{N}|x\rangle.
\end{equation}
It is invariant under every permutation. Moreover, the natural representation decomposes multiplicity-free as
\begin{equation}\label{eq:sec4_SN_decomposition}
\mathbb{C}^N=\operatorname{span}\{|\eta\rangle\}\oplus|\eta\rangle^\perp,
\end{equation}
where the two sectors have dimensions $1$ and $N-1$, respectively. The corresponding projectors are
\begin{equation}\label{eq:sec4_SN_projectors}
\Pi_1=|\eta\rangle\langle\eta|,
\qquad
\Pi_2=I-\Pi_1.
\end{equation}
To verify irreducibility of the second sector, let $W\subseteq|\eta\rangle^\perp$ be a nonzero invariant subspace and choose a nonzero vector $|z\rangle=\sum_xz_x|x\rangle\in W$. Then $\sum_xz_x=0$, so $z_i\ne z_j$ for some $i\ne j$. Invariance gives $(I-P_{ij})|z\rangle=(z_i-z_j)(|i\rangle-|j\rangle)\in W$. Applying further permutations shows that $W$ contains every $|p\rangle-|q\rangle$, and hence $W=|\eta\rangle^\perp$.

In Ref.~\cite{Zhou2024CohFraction}, the coherence fraction of the input state
is defined as
\begin{equation}\label{eq:sec4_fc}
f_c(|b\rangle):=|\langle\eta|b\rangle|^2.
\end{equation}
which can be rewritten as
$f_c(|b\rangle)=F(|b\rangle\langle b|,
|\eta\rangle\langle\eta|)^2$, where $F$ is the root fidelity.
In general, the coherence fraction is neither an entanglement measure
nor a coherence measure \cite{Zhou2024CohFraction}.
It describes the overlap with the equal-superposition state
fixed by the task.

\noindent {\bf Theorem 2}
For any nonsingular Hermitian matrix $A$, define the two projectors
$\Pi_1=|\eta\rangle\langle\eta|,
\Pi_2=I-\Pi_1,$
where
$|\eta\rangle=\frac{1}{\sqrt{N}}\sum_{x=1}^{N}|x\rangle .$
Then the permutation twirling of the success operator satisfies
\begin{equation}\label{eq:sec4_perm_twirled_operator}
\mathcal{T}_{S_N}(M_A)
=
\mu_1(A)\Pi_1+\mu_2(A)\Pi_2 ,
\end{equation}
and the success probability averaged over all permutation-generated right-hand sides satisfies
\begin{equation}\label{eq:sec4_perm_average_formula}
\overline{P}_{\mathrm{perm}}(|b\rangle;A)
=
\mu_2(A)
+
[\mu_1(A)-\mu_2(A)]f_c(|b\rangle).
\end{equation}
where the coefficients are given by
$
\mu_1(A)
:=
\gamma^2\langle\eta|A^{-2}|\eta\rangle ,
$
and
$
\mu_2(A)
:=
\frac{\operatorname{Tr}(M_A)-\mu_1(A)}
{N-1}.
$

The proof is given in Appendix~C. The two coefficients in Eq.~(\ref{eq:sec4_perm_twirled_operator})
are from the invariant direction $|\eta\rangle$ and its orthogonal
complement under the permutation representation. Consequently, the
affine law holds for arbitrary spectra of $A$ and $M_A$; it does not
require either operator to have only two distinct eigenvalues.
After averaging, the input dependence is completely captured by
the overlap $f_c(|b\rangle)$.

For convenience, define the permutation-sector contrast
\begin{equation*}
\Delta_{\mathrm{perm}}(A):=\mu_1(A)-\mu_2(A).
\end{equation*}
Then we obtain
\begin{equation}\label{eq:sec4_delta_perm_explicit}
\Delta_{\mathrm{perm}}(A)
=\frac{N\gamma^2}{N-1}
\left[
\langle\eta|A^{-2}|\eta\rangle
-\frac{1}{N}\operatorname{Tr}(A^{-2})
\right].
\end{equation}
If
$A=\sum_{j=1}^{N}\lambda_j|v_j\rangle\langle v_j|$
and
$w_j:=|\langle v_j|\eta\rangle|^2$, then
\begin{equation*}
\Delta_{\mathrm{perm}}(A)
=\frac{N\gamma^2}{N-1}
\left[
\sum_{j=1}^{N}\frac{w_j}{\lambda_j^2}
-\frac{1}{N}\sum_{j=1}^{N}\frac{1}{\lambda_j^2}
\right],
\qquad
\sum_{j=1}^{N}w_j=1.
\end{equation*}
Thus, the contrast depends not only on the eigenvalues of $A$, but also on the relative orientation of its eigenvectors and $|\eta\rangle$. Equivalently, Eq.~(\ref{eq:sec4_perm_average_formula}) can be written as
\begin{equation}\label{eq:sec4_perm_baseline}
\overline{P}_{\mathrm{perm}}(|b\rangle;A)
=\frac{1}{N}\operatorname{Tr}(M_A)
+\Delta_{\mathrm{perm}}(A)\left[f_c(|b\rangle)-\frac{1}{N}\right].
\end{equation}
Consequently, for a fixed matrix $A$, the permutation-averaged
success probability increases with $f_c$ when
$\Delta_{\mathrm{perm}}(A)>0$ and decreases with $f_c$ when
$\Delta_{\mathrm{perm}}(A)<0$.
When $\Delta_{\mathrm{perm}}(A)=0$, it is independent of the
input state and equals $\operatorname{Tr}(M_A)/N$.
Every computational-basis state has coherence fraction $1/N$ and therefore the same permutation-averaged success probability $\operatorname{Tr}(M_A)/N$. Equation~(\ref{eq:sec4_perm_baseline}) expresses the deviation from this incoherent-input baseline. If $A$ is diagonal in the computational basis, then $\langle\eta|A^{-2}|\eta\rangle=\operatorname{Tr}(A^{-2})/N$, so $\Delta_{\mathrm{perm}}(A)=0$. If $|\eta\rangle$ belongs to the maximal or minimal eigenspace of $M_A$, the contrast is nonnegative or nonpositive, respectively.

The preceding observation also gives a fixed-spectrum statement. Let
$m_j=\gamma^2/\lambda_j^2$, $m_{\min}=\min_jm_j$, $m_{\max}=\max_jm_j$ and
$\overline m=N^{-1}\sum_jm_j$. For the unitarily rotated family $A_U=UAU^\dagger$, one has
\begin{equation}\label{eq:sec4_fixed_spectrum_range}
\Delta_{\mathrm{perm}}(A_U)
\in
\left[
\frac{N}{N-1}(m_{\min}-\overline m),
\frac{N}{N-1}(m_{\max}-\overline m)
\right].
\end{equation}
Every value in this interval is attainable because $U^\dagger|\eta\rangle$ ranges over all normalized vectors, and the Rayleigh quotient of $M_A$ ranges over $[m_{\min},m_{\max}]$. If $M_A$ is not a scalar operator, then $m_{\min}<\bar{m}<m_{\max}$, so the interval contains negative, zero, and positive values. Thus, unitarily equivalent coefficient matrices can exhibit opposite responses to increasing coherence fraction, despite having identical eigenvalues and condition numbers. The sign of this response therefore depends on eigenvector alignment.

For a mixed input state $\rho$, define
$f_c(\rho):=\langle\eta|\rho|\eta\rangle$.
By linearity, Eq.~(\ref{eq:sec4_perm_average_formula}) extends to
\begin{equation*}
\overline{P}_{\mathrm{perm}}(\rho;A)
=\operatorname{Tr}\left[\rho\mathcal{T}_{S_N}(M_A)\right]
=\mu_2(A)+\Delta_{\mathrm{perm}}(A)f_c(\rho).
\end{equation*}
Every incoherent mixed state satisfies $f_c(\delta) = 1/N$ and therefore has permutation-averaged success probability $\operatorname{Tr}(M_A)/N$. This formula concerns the permutation average and does not determine the success probability of each individual input.
\vskip0.2in

\noindent {\bf 4.2 Numerical results}

We next present two numerical examples. The first example demonstrates positive, zero and negative permutation-sector contrasts for matrices with the same spectrum and condition number. The second example keeps the spectrum fixed and continuously rotates the eigenvectors to display the transition between these three regimes. All calculations use the ideal exact-spectral success operator $M_A=\gamma^2A^{-2}$ and explicitly average over all permutations.

Take
$
N=8,
\kappa=4,
\gamma=\frac{1}{4},
$
and fix the spectrum
\begin{equation}\label{eq:sec4_num_spectrum}
\Lambda=\operatorname{diag}\left(\frac14,\frac25,\frac12,\frac35,\frac7{10},\frac45,\frac9{10},1\right).
\end{equation}
Let $|q_1\rangle=|\eta\rangle$ and, for $j=2,\ldots,8$, define
\begin{equation}\label{eq:sec4_num_qj}
|q_j\rangle
=\frac{1}{\sqrt{j(j-1)}}
\left(\sum_{x=1}^{j-1}|x\rangle-(j-1)|j\rangle\right).
\end{equation}
The vectors $\{|q_j\rangle\}_{j=1}^{8}$ form an orthonormal basis. Let
$Q=(q_1,q_2,\ldots,q_8)$.

\smallskip

\noindent {\it Example 1. Positive, zero and negative permutation-sector contrasts.}
First, define
\begin{equation}\label{eq:sec4_num_Aplus}
A_+=Q\Lambda Q^T.
\end{equation}
Since $|\eta\rangle$ corresponds to the smallest eigenvalue $1/4$, it corresponds to the largest eigenvalue of $M_{A_+}$. We obtain
\begin{equation*}
\mu_1(A_+)=1,\qquad\mu_2(A_+)\simeq0.16844341,
\qquad\Delta_{\mathrm{perm}}(A_+)\simeq0.83155659>0,
\end{equation*}
and hence
$\overline{P}_{\mathrm{perm}}^{(+)}
\simeq0.16844341+0.83155659f_c.$
Second, take the computational-basis diagonal matrix
\begin{equation}
A_0=\Lambda.
\end{equation}
Since $A_0^{-2}$ is diagonal,
$\langle\eta|A_0^{-2}|\eta\rangle=N^{-1}\operatorname{Tr}(A_0^{-2})$ and therefore
$
\Delta_{\mathrm{perm}}(A_0)=0.
$
Thus,
$\overline{P}_{\mathrm{perm}}^{(0)}\simeq0.27238798.$
Finally, define
$
\Lambda_-=\operatorname{diag}\left(1,\frac25,\frac12,\frac35,\frac7{10},\frac45,\frac9{10},\frac14\right)
$
and
\begin{equation}\label{eq:sec4_num_Aminus}
A_-=Q\Lambda_-Q^T.
\end{equation}
\begin{figure}[ht]
\centering {\begin{minipage}[figure1]{0.8\linewidth}
\includegraphics[width=0.8\textwidth]{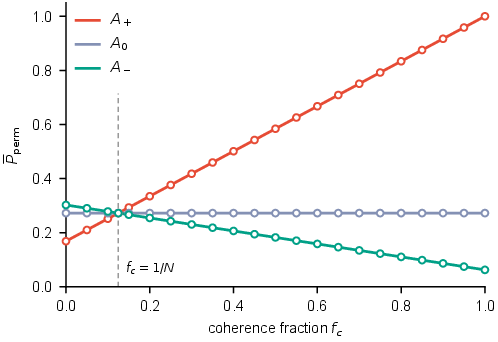}
\end{minipage}}
\caption{{Permutation-averaged success probability versus coherence fraction $f_c$ for the three matrices in Example 1. The red, gray-blue and green solid curves
correspond to $\overline{P}_{\rm perm}^{(+)}$, $\overline{P}_{\rm perm}^{(0)}$ and
$\overline{P}_{\rm perm}^{(-)}$, respectively, and the open circles denote the
corresponding numerical values.\label{fig:Example1}}}
\end{figure}
The three matrices have exactly the same spectrum and condition number. In this case, $|\eta\rangle$ corresponds to the largest eigenvalue $1$ of $A_-$ and hence to the smallest eigenvalue of $M_{A_-}$. We have
\begin{equation*}
\mu_1(A_-)=\frac{1}{16}=0.0625,\qquad\mu_2(A_-)\simeq0.30237198,\qquad
\Delta_{\mathrm{perm}}(A_-)\simeq-0.23987198<0,
\end{equation*}
and therefore
$\overline{P}_{\mathrm{perm}}^{(-)}\simeq0.30237198-0.23987198f_c.$
To vary the coherence fraction continuously, use the input family
\begin{equation}\label{eq:sec4_num_input_family}
|b(f)\rangle=\sqrt{f}|q_1\rangle+\sqrt{1-f}|q_2\rangle,
\qquad
0\le f\le1.
\end{equation}
Since $|q_1\rangle=|\eta\rangle$ and $\langle q_1|q_2\rangle=0$, one has
$f_c(|b(f)\rangle)=f$.
For $f=0,0.05,0.10,\ldots,1$, together with the additional point $f=1/8$, we evaluate
\begin{equation}\label{eq:sec4_num_average}
\overline{P}_{\mathrm{perm}}^{\mathrm{num}}(|b(f)\rangle;A)
=\frac{1}{8!}\sum_{\pi\in S_8}
\langle b(f)|P_\pi^\dagger M_AP_\pi|b(f)\rangle.
\end{equation}
The numerical averages agree with the analytical permutation-averaged expression, with a maximum absolute deviation below $10^{-14}$ in double precision.
Fig.~\ref{fig:Example1} compares the three affine relations with their direct permutation averages. Their common success-operator trace gives the intersection at $f_c = 1/8$, marked by the dashed line, where the averaged probability is approximately $0.27238798$.

\smallskip

\noindent {\it Example 2. A continuous change of direction at fixed spectrum.}
Keep the spectrum in Eq.~(\ref{eq:sec4_num_spectrum}) fixed and rotate only the eigenvectors associated with $1/4$ and $1$. Define
\begin{equation*}
\begin{aligned}
|v_{\min}(\theta)\rangle&=\cos\theta|q_1\rangle+\sin\theta|q_8\rangle,\\
|v_{\max}(\theta)\rangle&=-\sin\theta|q_1\rangle+\cos\theta|q_8\rangle,
\end{aligned}
\qquad
0\le\theta\le\frac{\pi}{2}.
\end{equation*}
Let
$Q(\theta)=(v_{\min}(\theta),q_2,\ldots,q_7,v_{\max}(\theta))$
and define
\begin{equation}\label{eq:sec4_num_rotated_matrix}
A(\theta)=Q(\theta)\Lambda Q(\theta)^T.
\end{equation}
For all $\theta$, $A(\theta)$ has the same spectrum and condition number as $A_+$. Direct calculation gives
$\mu_1(\theta)=\cos^2\theta+\frac{1}{16}\sin^2\theta.$
Let
$S:=\operatorname{Tr}(M_{A(\theta)})
=\gamma^2\operatorname{Tr}(\Lambda^{-2})
\simeq2.17910387534643.$
Since $S$ is independent of $\theta$,
\begin{equation*}
\mu_2(\theta)=\frac{S-\mu_1(\theta)}{7},
\qquad
\Delta_{\mathrm{perm}}(\theta)=\frac{8\mu_1(\theta)-S}{7}.
\end{equation*}
The transition angle determined by $\Delta_{\mathrm{perm}}(\theta_0)=0$ is
$\cos^2\theta_0=\frac{S/8-1/16}{1-1/16},$
$\theta_0\simeq61.76037572^\circ.$
At this angle,
$\mu_1(\theta_0)=\mu_2(\theta_0)=S/8\simeq0.27238798$, so the permutation-averaged success probability is independent of $f_c$.

\begin{figure}[ht]
\centering {\begin{minipage}[figure2]{0.8\linewidth}
\includegraphics[width=0.8\textwidth]{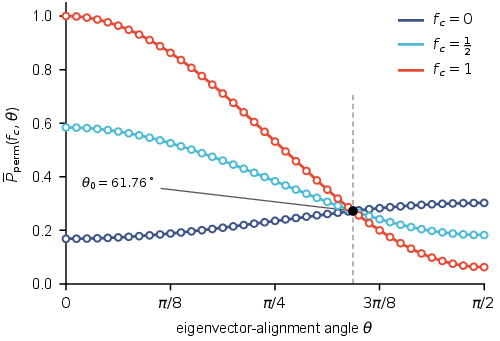}
\end{minipage}}
\caption{{Permutation-averaged success probability versus
eigenvector-alignment angle $\theta$.
The blue, cyan and red solid curves denote the analytical values for
$f_c=0$, $1/2$ and $1$, respectively, and the open circles denote the
corresponding numerical values.
\label{fig:Example2}}}
\end{figure}

We evaluate Eq.~(\ref{eq:sec4_num_average}) for the three input values $f_c=0,1/2,1$ over the range $\theta\in[0,\pi/2]$. The corresponding analytical expression is
$\overline{P}_{\mathrm{perm}}(f_c,\theta)=\mu_2(\theta)+\Delta_{\mathrm{perm}}(\theta)f_c .$ The numerical averages agree with this analytical expression, with a maximum absolute deviation below $10^{-14}$.
Fig.~\ref{fig:Example2} shows that larger coherence fractions yield higher averaged success probabilities for $ \theta < \theta_0 $, whereas the opposite holds for $ \theta > \theta_0 $. The three curves intersect at $ \theta = \theta_0 $. The endpoint matrices $ A(0) = A_+ $ and $ A(\pi/2) = A_- $ recover the opposite-contrast cases of Example 1.

\vskip0.2in

\noindent {\bf 5. Structured-target preprocessed WZP algorithm}

Unlike the ensemble average in Section 4, we now optimize over the allowed preprocessing operations. Once the success operator has only two spectral levels, the restricted input problem admits an exact affine reduction to a projector-overlap optimization. This motivates us to investigate how admissible preprocessing affects the success probability of the WZP algorithm in the structured-target setting.

\vskip0.2in

\noindent {\bf 5.1 Preprocessed WZP algorithm}

The motivation for introducing preprocessing is to investigate whether the input state can be better aligned with the favorable spectral subspace of the success operator before the WZP algorithm is applied. As shown in Section 3, the success probability is governed by the overlap between the input state and the dominant spectral subspaces of the success operator. This naturally raises the question of whether such an overlap, and hence the success probability, can be increased by applying an admissible transformation to the input state.
To address this question, we introduce the preprocessed WZP algorithm, in which a preprocessing unitary $V$ chosen from a prescribed admissible subgroup $\mathcal{G} \subseteq U(N)$ is applied to the input state before the WZP algorithm. The resulting success probability is therefore $P_0(V|b\rangle)$, and optimizing over $V \in \mathcal{G}$ quantifies the maximal success probability achievable under the prescribed preprocessing restriction.

It should be emphasized that the preprocessing unitary acts on the input state before the WZP algorithm. Unless the coefficient matrix is transformed consistently as well, this operation changes the right-hand side and therefore generates a family of related linear systems rather than an equivalent representation of a single fixed linear system.
Consequently, the role of preprocessing considered here is to characterize the maximal postselection success probability achievable under admissible transformations of the input state.

\begin{figure}[ht]
\centering {\begin{minipage}[figure3]{1.0\linewidth}
\includegraphics[width=1.0\textwidth]{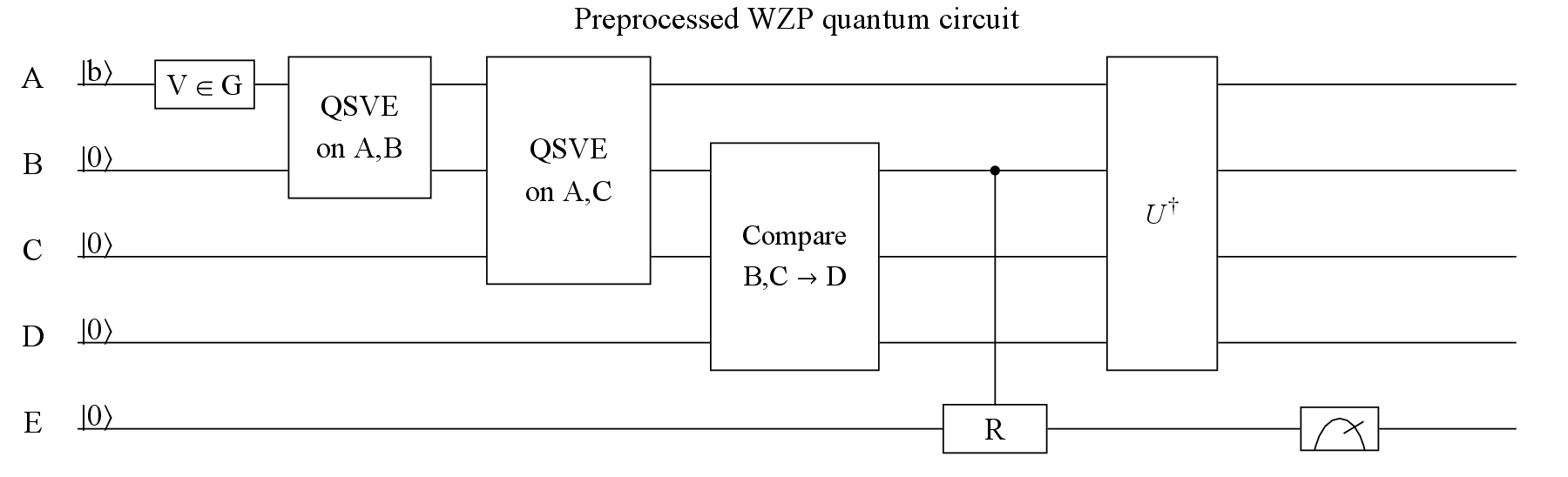}
\end{minipage}}
\caption{{Quantum circuit for the preprocessed WZP algorithm.\label{fig:Fig3}}}
\end{figure}

As shown in Fig.~\ref{fig:Fig3}, the preprocessed WZP algorithm consists of an admissible preprocessing unitary $V \in \mathcal{G}$ followed by the WZP algorithm.
Indeed, any positive definite Hermitian operator with exactly two eigenvalues $0<a<b$ can be written as
$A=a\Pi+b(I-\Pi)=
b\Bigl(\frac{a}{b}\Pi+(I-\Pi)\Bigr),$
for a nonzero proper orthogonal projector $\Pi$. Thus, after normalizing the two eigenvalue levels by their ratio $k:=b/a>1$, it suffices to study the family
\begin{equation}\label{eq:sec5_AkPi}
A_{k,\Pi}
=
\frac{1}{k}\Pi+(I-\Pi),
\qquad
k>1.
\end{equation}
For \(A_{k,\Pi}\), the two eigenvalues are \(1/k\) and \(1\), so its condition number is \(k\).
Thus \(k\) coincides with the condition-number parameter \(\kappa\) used in the general WZP discussion. Throughout this section, we set $\gamma=1/k$.

To characterize the effect of preprocessing in this two-level setting, we consider an arbitrary normalized pure input state $|b\rangle \in \mathbb{C}^N$ and an admissible compact preprocessing subgroup $\mathcal{G} \subseteq U(N)$. Rather than restricting the target to a rank-one state, we allow a general rank-$d$ target subspace, with $1 \leq d < N$. Let $\Pi$ be the orthogonal projector onto this target subspace. Then
\begin{equation}\label{eq:sec5_proj}
\Pi^2=\Pi,
\qquad
\Pi^\dagger=\Pi,
\qquad
\operatorname{rank}(\Pi)=d.
\end{equation}
With $\gamma = 1/k$, the success operator associated with $A_{k,\Pi}$ takes the particularly simple form
\begin{equation*}
M = \gamma^2 A_{k,\Pi}^{-2}
= \Pi + \frac{1}{k^2}(I-\Pi).
\end{equation*}
Denoting by $\operatorname{Ran}(X)$ the range of a linear operator $X$, $\operatorname{Ran}(\Pi)$ is the target subspace and, from the above expression for $M$, also the dominant spectral subspace of the success operator, with eigenvalue $1$, while its orthogonal complement has eigenvalue $1/k^2$. Therefore, increasing the success probability amounts to increasing the projection weight of the preprocessed input state onto $\operatorname{Ran}(\Pi)$. In this way, the effect of preprocessing is reduced to a geometric alignment problem between the input state and the target subspace.

\vskip0.2in

\noindent {\bf 5.2 Success probability under subgroup-constrained preprocessing}

We now characterize the optimal success probability of the preprocessed WZP algorithm under the admissible preprocessing subgroup $\mathcal{G}$. Define the projector orbit
$
\mathcal{O}_{\mathcal{G}}(\Pi)
:=
\{V^\dagger \Pi V : V \in \mathcal{G}\},
$
the orbit-overlap functional
$
\Omega_{\mathcal{G},\Pi}(|b\rangle)
:=
\max_{V\in\mathcal{G}}
\langle b|V^\dagger \Pi V|b\rangle,
$
and the optimal success probability of the preprocessed WZP algorithm
\begin{equation*}
P_{0,\max}^{(\Pi,\mathcal{G})}(|b\rangle)
:=
\max_{V\in\mathcal{G}} P_0(V|b\rangle).
\end{equation*}
\noindent {\bf Proposition 1}
For every normalized pure state $|b\rangle$ and every $V\in\mathcal{G}$,
$P_0(V|b\rangle)=\frac{1}{k^2}+
\left(1-\frac{1}{k^2}\right)\|\Pi V|b\rangle\|^2.$
Consequently,
\begin{equation}\label{eq:sec5_prop1_exact}
P_{0,\max}^{(\Pi,\mathcal{G})}(|b\rangle)
=
\frac{1}{k^2}
+
\left(1-\frac{1}{k^2}\right)\Omega_{\mathcal{G},\Pi}(|b\rangle).
\end{equation}
$\it{Proof}$.
Since $\Pi$ is an orthogonal projector, we have
$A_{k,\Pi}^{-2}=k^2\Pi+(I-\Pi).$
Using $\gamma=1/k$, we obtain
\begin{equation*}
M_{k,\Pi}
=
\gamma^2 A_{k,\Pi}^{-2}
=
\frac{1}{k^2}I+\left(1-\frac{1}{k^2}\right)\Pi.
\end{equation*}
Now let $V\in\mathcal{G}\subseteq U(N)$. Then
\begin{align*}
P_0(V|b\rangle)
=
\langle b|V^\dagger M_{k,\Pi}V|b\rangle=
\frac{1}{k^2}
+
\left(1-\frac{1}{k^2}\right)\langle Vb|\Pi|Vb\rangle.
\end{align*}
Since $\Pi$ is an orthogonal projector, $\langle Vb|\Pi|Vb\rangle=\|\Pi V|b\rangle\|^2$. Maximizing over $V\in\mathcal{G}$ yields Eq.~(\ref{eq:sec5_prop1_exact}). \qed

The formulas below extend to any two-level success operator
$M = \omega_2 I + (\omega_1 - \omega_2)\Pi$,
with $0 < \omega_2 < \omega_1 \leq 1$,
by replacing $k^{-2}$ with $\omega_2$ and
$1-k^{-2}$ with $\omega_1-\omega_2$.
If $\mathcal{G}=U(N)$, any normalized input can be mapped into
the nonzero target subspace, giving
$\Omega_{\mathcal{G},\Pi}(|b\rangle) = 1$.
The optimum then equals $\omega_1$, which is $1$
for $A_{k,\Pi}$.
Restrictions on $\mathcal{G}$ are therefore essential for
nontrivial input dependence.

Let $\Pi_d^{\mathrm{inc}}=\sum_{j=1}^{d}|j\rangle\langle j|$
be the canonical rank-$d$ coordinate projector. The natural subgroup that preserves the incoherent pure-state set (up to phases) is the group of incoherent unitaries \cite{Chitambar2016},
\begin{equation*}
\mathcal{G}_{\mathrm{inc}}
=
\left\{
DP:\ D=\operatorname{diag}(e^{\mathrm{i}\theta_1},\dots,e^{\mathrm{i}\theta_N}),\ P\text{ a permutation matrix}
\right\}.
\end{equation*}
Write the input state as
$|b\rangle=\sum_{x=1}^{N} b_x |x\rangle$,
$\sum_{x=1}^{N}|b_x|^2=1,$
define
$p_x:=|b_x|^2,$
and let
$
p_1^\downarrow\ge p_2^\downarrow\ge\cdots\ge p_N^\downarrow
$
be the nonincreasing rearrangement of $\{p_x\}_{x=1}^{N}$. We have (see proof in Appendix D)

\smallskip

\noindent {\bf Theorem 3} The maximal success probability is given by
\begin{equation}\label{eq:sec5_thm3_P}
P_{0,\max}^{(\Pi_d^{\mathrm{inc}},\mathcal{G}_{\mathrm{inc}})}(|b\rangle)
=
\frac{1}{k^2}
+
\left(1-\frac{1}{k^2}\right)\sum_{j=1}^{d} p_j^\downarrow.
\end{equation}
In particular, when $d=1$,
$P_{0,\max}^{(\Pi_1^{\mathrm{inc}},\mathcal{G}_{\mathrm{inc}})}(|b\rangle)
=
\frac{1}{k^2}
+
\left(1-\frac{1}{k^2}\right)\bigl(1-C_g(|b\rangle)\bigr),$
where $C_g$ is the geometric coherence of the pure state $|b\rangle$.

Denote $\mathcal{H}=\mathcal{H}_A\otimes\mathcal{H}_B,$
with $\dim(\mathcal{H}_A)=N_A,$ $\dim(\mathcal{H}_B)=N_B$ and $N=N_A N_B.$
Fix an integer $1\le r\le \min\{N_A,N_B\}$, we define
$\Pi_r^{\mathrm{prod}}
:=\sum_{j=1}^{r}|jj\rangle\langle jj|.$
The admissible preprocessing subgroup is the local unitary group,
$\mathcal{G}_{A|B}=U(N_A)\otimes U(N_B)$.
Let $\mathrm{SR}(|\phi\rangle)$ be the Schmidt rank
of a bipartite pure state $|\phi\rangle\in\mathcal{H}_A\otimes\mathcal{H}_B$.
Define the rank-$r$ geometric entanglement with respect to the bipartition $A|B$ by
\[
E_{g,r}^{A|B}(|b\rangle)
:=
1-\max_{\mathrm{SR}(|\phi\rangle)\le r} |\langle \phi|b\rangle|^2.
\]
Let $\mu_1\ge \mu_2\ge\cdots\ge 0$ be the eigenvalues of the reduced state $\rho_A=\operatorname{Tr}_B(|b\rangle\langle b|)$ in nonincreasing order.
We have the following result, see proof in Appendix E.

\smallskip

\noindent {\bf Theorem 4}
The maximal success probability is given by
\begin{equation}\label{eq:sec5_thm4_P}
P_{0,\max}^{(\Pi_r^{\mathrm{prod}},\mathcal{G}_{A|B})}(|b\rangle)
=
\frac{1}{k^2}
+
\left(1-\frac{1}{k^2}\right)\bigl(1-E_{g,r}^{A|B}(|b\rangle)\bigr).
\end{equation}
In particular, when $r=1$,
$P_{0,\max}^{(\Pi_1^{\mathrm{prod}},\mathcal{G}_{A|B})}(|b\rangle)
=
\frac{1}{k^2}
+
\left(1-\frac{1}{k^2}\right)\bigl(1-E_g^{A|B}(|b\rangle)\bigr)$.

For this target subspace, the maximal projection weight attainable
by local unitaries equals the largest squared overlap of $|b\rangle$
with a normalized state of Schmidt rank at most $r$.
This directly links the optimal success probability
to rank-$r$ geometric entanglement.

Furthermore, we have the following corollaries.

\noindent {\bf Corollary 2 }
With respect to the rank-one projector $\Pi_{\mathbb{R}}:=|1\rangle\langle 1|$,
if the admissible preprocessing subgroup is the real orthogonal group $O(N)\subset U(N)$, then
\begin{equation}\label{eq:sec5_imag}
P_{0,\max}^{(\Pi_{\mathbb{R}},O(N))}(|b\rangle)
=
\frac{1}{k^2}
+
\left(1-\frac{1}{k^2}\right)\bigl(1-I_g(|b\rangle)\bigr).
\end{equation}
$\it{Proof}$.
For any $O\in O(N)$,
$O^\dagger \Pi_{\mathbb{R}} O=
O^T |1\rangle\langle 1| O=
|r_O\rangle\langle r_O|,
|r_O\rangle:=O^T|1\rangle.$
As $O$ ranges over $O(N)$, the vector $|r_O\rangle$ ranges over all real unit vectors. Therefore,
\[
\Omega_{O(N),\Pi_{\mathbb{R}}}(|b\rangle)
=
\max_{|r\rangle\in\mathcal{R}} |\langle r|b\rangle|^2
=
1-I_g(|b\rangle).
\]
Applying Proposition~1, one obtains Eq.~(\ref{eq:sec5_imag}). \qed

\noindent {\bf Corollary 3} Assume
$N=2^n$. With respect to any fully separable pure state
$|\phi_{\mathrm{sep}}\rangle=\bigotimes_{j=1}^n |\alpha_j\rangle,$
if the admissible preprocessing subgroup is the local unitary group
$
G_{\mathrm{loc}}:=U(2)^{\otimes n},
$
then
\[
P^{(\Pi_{\mathrm{sep}},G_{\mathrm{loc}})}_{0,\max}(|b\rangle)
=
\frac{1}{k^2}
+
\left(1-\frac{1}{k^2}\right)\bigl(1-E_g(|b\rangle)\bigr),
\]
where $\Pi_{\mathrm{sep}}=|\phi_{\mathrm{sep}}\rangle\langle\phi_{\mathrm{sep}}|$.\\
$\it{Proof}$.
The orbit of $\Pi_{\mathrm{sep}}$ under $G_{\mathrm{loc}}$ is exactly the set of
rank-one projectors onto fully separable pure states. Hence
\[
\Omega_{G_{\mathrm{loc}},\Pi_{\mathrm{sep}}}(|b\rangle)
=
\max_{|\phi\rangle\in S_p} |\langle \phi|b\rangle|^2
=
1-E_g(|b\rangle).
\]
Applying Proposition 1, one proves the claim. \qed

For the rank-one targets considered above, Corollaries 2 and 3 relate the optimal success probability to geometric imaginarity and multipartite geometric entanglement under real orthogonal and local-unitary preprocessing, respectively. Together with Theorems 3 and 4, these results show that the relevant geometric quantity is determined jointly by the admissible transformations and the target subspace. Within the two-level family, the maximal accessible overlap then determines the optimal postselection success probability exactly.

\vskip0.1in

\noindent {\bf 6. Conclusions}\\\hspace*{\fill}\\
In this work, we have extended the original WZP algorithm from real symmetric linear systems to complex Hermitian linear systems and studied the success probability of the modified WZP algorithm from a resource-theoretic perspective. We expressed the success probability as the expectation value of a positive operator. For restricted input sets, we derived general bounds in terms of the spectral structure of the success operator and the maximal overlap with its dominant spectral subspaces.
In the nondegenerate leading-eigenvalue case, these bounds
are characterized by geometric resource quantities, including
geometric entanglement, geometric coherence and geometric
imaginarity.

We have also considered the success probability of the modified WZP algorithm averaged over permutation-generated right-hand sides. In this setting, the averaged success probability is exactly an affine function of the coherence fraction, with the slope determined by the alignment of the success operator with the equal-superposition state. In particular, the same spectrum and condition number may lead to positive, zero, or negative dependence on the coherence fraction. This shows that the relation between the coherence fraction and the permutation-averaged success probability is intrinsically task-dependent and cannot be determined by spectral information alone.

For the two-level family, the optimal success probability is determined by the maximal projection weight onto the orbit of the target projector. For different preprocessing subgroups, this relation yields exact characterizations in terms of geometric coherence, geometric entanglement and geometric imaginarity. Overall, our results provide a unified geometric perspective on the WZP success probability under input restrictions, permutation averaging, and constrained preprocessing. They highlight how geometric resource measures characterize the success probability beyond spectral properties alone and may provide a useful basis for studying other quantum algorithms with resource and symmetry constraints.

\subsubsection*{Acknowledgements}
\small {This work was supported by National Natural Science
Foundation of China (Grant Nos. 12561084, 12161056); Natural Science Foundation of Jiangxi Province (Grant No. 20232ACB211003).}

\begin{center}
\textbf{Appendix A. Proof of Theorem 1}
\end{center}
Let $|b\rangle$ be any normalized pure state. Define
$q_\ell=\langle b|\Lambda_\ell|b\rangle,$
$q_\ell\ge 0,$
$\sum_{\ell=1}^{L} q_\ell=1.$
Set
$s_\ell=\sum_{t=1}^{\ell} q_t=
\langle b|\Pi_\ell|b\rangle,$
$\ell=1,\dots,L,$
with $s_0=0$. Then $q_\ell=s_\ell-s_{\ell-1}$, and
\begin{align*}
P_0(|b\rangle)
=
\sum_{\ell=1}^{L}\omega_\ell q_\ell
=
\omega_L+\sum_{\ell=1}^{L-1}(\omega_\ell-\omega_{\ell+1})s_\ell.
\end{align*}
Since $s_\ell\le \Gamma_S(\ell)$ for every admissible $|b\rangle\in S$, we have
\begin{equation*}
P_{\max}^S
\le
\omega_L+\sum_{\ell=1}^{L-1}(\omega_\ell-\omega_{\ell+1})\Gamma_S(\ell),
\end{equation*}
which proves the upper bound in Eq.~(9).

For the lower bound, fix $\ell\in\{1,\dots,L\}$. Then
\begin{align*}
P_0(|b\rangle)
=
\sum_{t=1}^{\ell}\omega_t q_t+\sum_{t=\ell+1}^{L}\omega_t q_t
\ge
\omega_\ell \sum_{t=1}^{\ell} q_t
+
\omega_L \sum_{t=\ell+1}^{L} q_t
=
\omega_L+(\omega_\ell-\omega_L)s_\ell.
\end{align*}
Assume that $|b_\ell^\ast\rangle\in S$ satisfies
$
\langle b_\ell^\ast|\Pi_\ell|b_\ell^\ast\rangle=\Gamma_S(\ell).
$
Then
$
P_{\max}^S
\ge
P_0(|b_\ell^\ast\rangle)
\ge
\omega_L+(\omega_\ell-\omega_L)\Gamma_S(\ell).
$
Taking the maximum over $\ell$ proves the lower bound in Eq.~(9).
We next characterize the equality condition for the upper bound. Define
\[
U
:=
\omega_L
+
\sum_{\ell=1}^{L-1}
(\omega_{\ell}-\omega_{\ell+1})\Gamma_S(\ell).
\]
For every $|b\rangle\in S$, we have
\[
\begin{aligned}
U-P_0(|b\rangle)
&=
\sum_{\ell=1}^{L-1}
(\omega_{\ell}-\omega_{\ell+1})
\left[
\Gamma_S(\ell)-s_{\ell}(b)
\right]
\geq 0.
\end{aligned}
\]
Since $\omega_{\ell}-\omega_{\ell+1}>0$ for all
$\ell=1,\ldots,L-1$, equality holds if and only if
\[
s_{\ell}(b)=\Gamma_S(\ell),
\qquad
\ell=1,\ldots,L-1.
\]
Therefore, the upper bound in Eq.~(9) is attained if and only if
there exists an admissible state $|b_{*}\rangle\in S$ such that
\[
\langle b_{*}|\Pi_{\ell}|b_{*}\rangle
=
\Gamma_S(\ell),
\qquad
\ell=1,\ldots,L-1.
\]
Finally, fix
\[
\ell^{*}
\in
\arg\max_{1\leq \ell\leq L}
\left[
\omega_L
+
(\omega_{\ell}-\omega_L)\Gamma_S(\ell)
\right],
\]
and define
\[
B
:=
\omega_L
+
(\omega_{\ell^{*}}-\omega_L)\Gamma_S(\ell^{*}).
\]
For every $|b\rangle\in S$, we can write
\[
P_0(|b\rangle)
=
\omega_L
+
(\omega_{\ell^{*}}-\omega_L)s_{\ell^{*}}(b)
+
D_{\ell^{*}}(b),
\]
where
\[
D_{\ell^{*}}(b)
=
\sum_{t<\ell^{*}}
(\omega_t-\omega_{\ell^{*}})q_t(b)
+
\sum_{\ell^{*}<t<L}
(\omega_t-\omega_L)q_t(b).
\]
Hence,
\[
\begin{aligned}
P_0(|b\rangle)-B
&=
D_{\ell^{*}}(b)
-
(\omega_{\ell^{*}}-\omega_L)
\left[
\Gamma_S(\ell^{*})-s_{\ell^{*}}(b)
\right].
\end{aligned}
\]
Since the lower-bound argument above gives $P_{\max}^{S}\geq B$,
the lower bound is attained, namely $P_{\max}^{S}=B$, if and only if
\[
P_0(|b\rangle)\leq B
\]
for every $|b\rangle\in S$. By the preceding identity, this is
equivalent to
\[
D_{\ell^{*}}(b)
\leq
(\omega_{\ell^{*}}-\omega_L)
\left[
\Gamma_S(\ell^{*})-s_{\ell^{*}}(b)
\right],
\qquad
\forall\,|b\rangle\in S.
\]
This completes the proof.\qed

\begin{center}
\textbf{Appendix B. Proof of Lemma 1}
\end{center}
Write $X=\sum_{\alpha,\beta}X_{\alpha\beta}$, where
$X_{\alpha\beta}=\Pi_\alpha X\Pi_\beta$.
For every $h\in G$, Haar invariance implies
\begin{equation*}
U_h^\dagger\mathcal{T}_G(X)U_h=\mathcal{T}_G(X).
\end{equation*}
Therefore, each block of the twirled operator satisfies
\begin{equation*}
U_h^{(\alpha)\dagger}
\mathcal{T}_G(X)_{\alpha\beta}
U_h^{(\beta)}
=
\mathcal{T}_G(X)_{\alpha\beta}.
\end{equation*}
By Schur's lemma\cite{Bartlett2007}, the off-diagonal blocks between
inequivalent irreducible sectors vanish. Hence,
$\mathcal{T}_G(X)_{\alpha\beta}=0$ for $\alpha\neq\beta$.
For the diagonal blocks, Schur's lemma implies that
\begin{equation*}
\mathcal{T}_G(X_{\alpha\alpha})
=
Y_\alpha\otimes I_{\mathcal{H}_\alpha},
\end{equation*}
where $Y_\alpha$ acts on the multiplicity space.
Taking the partial trace over $\mathcal{H}_\alpha$ gives
\begin{equation*}
d_\alpha Y_\alpha
=
\operatorname{Tr}_{\mathcal{H}_\alpha}
(X_{\alpha\alpha}),
\end{equation*}
which determines $Y_\alpha$ and proves Eq.~(\ref{eq:sec4_general_twirling}).
Taking $X=M_A$, define the operators on the multiplicity spaces as
\begin{equation*}
K_\alpha(A):=\frac{1}{d_\alpha}
\operatorname{Tr}_{\mathcal{H}_\alpha}
\left(\Pi_\alpha M_A\Pi_\alpha\right),
~~~~\rho_\alpha^{(b)}:=
\operatorname{Tr}_{\mathcal{H}_\alpha}
\left(\Pi_\alpha |b\rangle\langle b|\Pi_\alpha\right).
\end{equation*}
Using the block decomposition of $\mathcal{T}_G(M_A)$ and evaluating the
expectation value in $|b\rangle$, we obtain
\begin{equation*}
\overline{P_G(|b\rangle;A)}
=
\sum_\alpha
\operatorname{Tr}
\left[
K_\alpha(A)\rho_\alpha^{(b)}
\right],
\end{equation*}
which proves Eq.~(\ref{eq:sec4_general_success}).
\qed

\begin{center}
\textbf{Appendix C. Proof of Theorem 2}
\end{center}
The natural permutation representation of $S_N$ decomposes into two inequivalent multiplicity-free irreducible sectors,
\begin{equation*}
\mathbb{C}^{N}
=
\operatorname{span}\{|\eta\rangle\}
\oplus
|\eta\rangle^\perp .
\end{equation*}
The corresponding projectors are
$
\Pi_1=|\eta\rangle\langle\eta|,
\Pi_2=I-\Pi_1 .$
Since the representation is multiplicity-free, applying Lemma 1 to $M_A$ gives
\begin{equation*}
\mathcal{T}_{S_N}(M_A)
=
c_1\Pi_1+c_2\Pi_2 .
\end{equation*}
For the one-dimensional sector,
$c_1
=
\operatorname{Tr}(\Pi_1M_A)
=
\langle\eta|M_A|\eta\rangle
=
\mu_1(A).$
Moreover, the trace is preserved under twirling, and hence
\begin{equation*}
\operatorname{Tr}(M_A)
=
\operatorname{Tr}(\mathcal{T}_{S_N}(M_A))
=
\mu_1(A)+(N-1)\mu_2(A).
\end{equation*}
Therefore $\mu_2(A)
=
\frac{\operatorname{Tr}(M_A)-\mu_1(A)}
{N-1}.
$
Then we have
\begin{equation*}
\mathcal{T}_{S_N}(M_A)
=
\mu_1(A)\Pi_1+\mu_2(A)\Pi_2 .
\end{equation*}
Finally, the permutation-averaged success probability is
\begin{equation*}
\overline{P}_{\mathrm{perm}}(|b\rangle;A)
=
\langle b|
\mathcal{T}_{S_N}(M_A)
|b\rangle .
\end{equation*}
Substituting the above decomposition gives
\begin{equation*}
\overline{P}_{\mathrm{perm}}(|b\rangle;A)
=
\mu_1(A)\langle b|\Pi_1|b\rangle
+
\mu_2(A)\langle b|\Pi_2|b\rangle .
\end{equation*}
Because
$
\langle b|\Pi_1|b\rangle
=
|\langle\eta|b\rangle|^2
=
f_c(|b\rangle),
$
and
$
\langle b|\Pi_2|b\rangle
=
1-f_c(|b\rangle),
$
we obtain
\begin{equation*}
\overline{P}_{\mathrm{perm}}(|b\rangle;A)
=
\mu_1(A)f_c(|b\rangle)
+
\mu_2(A)(1-f_c(|b\rangle)),
\end{equation*}
which completes the proof.\qed

\begin{center}
\textbf{Appendix D. Proof of Theorem 3}
\end{center}
Let $V=DP\in\mathcal{G}_{\mathrm{inc}}$. Since both $D$ and $\Pi_d^{\mathrm{inc}}$ are diagonal, $V^\dagger \Pi_d^{\mathrm{inc}} V=P^\dagger \Pi_d^{\mathrm{inc}} P.$
If $S_P:=\{P^{-1}(1),P^{-1}(2),\dots,P^{-1}(d)\},$
then $|S_P|=d$ and
\begin{equation*}
P^\dagger \Pi_d^{\mathrm{inc}} P
=
\sum_{x\in S_P}|x\rangle\langle x|
=: \Pi_{S_P}.
\end{equation*}
Conversely, every subset $S\subseteq\{1,\dots,N\}$ with $|S|=d$ can be realized in this way by a suitable permutation matrix. Hence
\begin{equation*}
\mathcal{O}_{\mathcal{G}_{\mathrm{inc}}}(\Pi_d^{\mathrm{inc}})
=
\left\{
\Pi_S=\sum_{x\in S}|x\rangle\langle x|:\ |S|=d
\right\}.
\end{equation*}
By Proposition~1,
\begin{align*}
\Omega_{\mathcal{G}_{\mathrm{inc}},\Pi_d^{\mathrm{inc}}}(|b\rangle)
=\max_{Q\in\mathcal{O}_{\mathcal{G}_{\mathrm{inc}}}(\Pi_d^{\mathrm{inc}})}
\langle b|Q|b\rangle
=\max_{|S|=d}\langle b|\Pi_S|b\rangle
=\max_{|S|=d}\sum_{x\in S}p_x.
\end{align*}
The maximum is attained when \(S\) is the set of indices of the \(d\) largest probabilities, yielding
\[
\Omega_{\mathcal{G}_{\mathrm{inc}},\Pi_d^{\mathrm{inc}}}(|b\rangle)
=
\sum_{j=1}^{d} p_j^\downarrow.
\]
Eq.~(\ref{eq:sec5_thm3_P}) then follows from Proposition~1.
When $d=1$,
$\sum_{j=1}^{1} p_j^\downarrow=\max_x |b_x|^2.$
For pure states, the geometric coherence satisfies
$
1-C_g(|b\rangle)=\max_x |b_x|^2.
$
Substituting this into Eq.~(\ref{eq:sec5_thm3_P}) proves $P_{0,\max}^{(\Pi_1^{\mathrm{inc}},\mathcal{G}_{\mathrm{inc}})}(|b\rangle)
=
\frac{1}{k^2}
+
\left(1-\frac{1}{k^2}\right)\bigl(1-C_g(|b\rangle)\bigr)$. \qed

\begin{center}
\textbf{Appendix E. Proof of Theorem 4}
\end{center}
Let $V=U_A\otimes U_B\in\mathcal{G}_{A|B}$. Then
$V^\dagger \Pi_r^{\mathrm{prod}} V
=
\sum_{j=1}^{r}|a_j b_j\rangle\langle a_j b_j|,$
where $\{a_j\}_{j=1}^{r}\subset\mathcal{H}_A$ and
$\{b_j\}_{j=1}^{r}\subset\mathcal{H}_B$ are orthonormal sets. Hence the orbit
$\mathcal{O}_{\mathcal{G}_{A|B}}(\Pi_r^{\mathrm{prod}})$ consists exactly of projectors of the form
\begin{equation*}
Q=\sum_{j=1}^{r}|a_j b_j\rangle\langle a_j b_j|.
\end{equation*}
Every normalized vector in $\operatorname{Ran}(Q)$ is a linear combination of
$r$ product vectors and therefore has Schmidt rank at most $r$. Thus
\begin{equation*}
\langle b|Q|b\rangle
=
\|Q|b\rangle\|^2
=
\max_{\substack{|\phi\rangle\in\operatorname{Ran}(Q)\\ \|\phi\|=1}}
|\langle \phi|b\rangle|^2
\le
\max_{\operatorname{SR}(|\phi\rangle)\le r}
|\langle \phi|b\rangle|^2.
\end{equation*}
Conversely, if $|\phi\rangle$ has Schmidt rank at most $r$, then it admits a Schmidt
decomposition
$|\phi\rangle=\sum_{j=1}^{r}s_j|a_j b_j\rangle$
for suitable orthonormal sets $\{a_j\}$ and $\{b_j\}$. Hence
$|\phi\rangle\in\operatorname{Ran}(Q)$ for some orbit projector $Q$, and therefore
\begin{equation*}
|\langle \phi|b\rangle|^2
=
|\langle \phi|Q|b\rangle|^2
\le
\|Q|b\rangle\|^2
=
\langle b|Q|b\rangle
\le
\Omega_{\mathcal{G}_{A|B},\Pi_r^{\mathrm{prod}}}(|b\rangle).
\end{equation*}
Taking the maximum over all such $|\phi\rangle$, we obtain
$\Omega_{\mathcal{G}_{A|B},\Pi_r^{\mathrm{prod}}}(|b\rangle)
=
\max_{\operatorname{SR}(|\phi\rangle)\le r}
|\langle \phi|b\rangle|^2$.

Let $|b\rangle=\sum_{j=1}^{R}\sqrt{\mu_j}\,|\alpha_j\beta_j\rangle$
be a Schmidt decomposition of $|b\rangle$, where
$\mu_1\ge \mu_2\ge \cdots\ge \mu_R \ge 0$ are the eigenvalues of $\rho_A$.
Choosing
$Q_\star
=
\sum_{j=1}^{r}|\alpha_j\beta_j\rangle\langle \alpha_j\beta_j|$
gives
\begin{equation*}
\Omega_{\mathcal{G}_{A|B},\Pi_r^{\mathrm{prod}}}(|b\rangle)
\ge
\langle b|Q_\star|b\rangle
=
\sum_{j=1}^{r}\mu_j.
\end{equation*}
For the opposite inequality, let
$Q=\sum_{j=1}^{r}|a_j b_j\rangle\langle a_j b_j|$
be any orbit projector, and define
$Q_A:=\sum_{j=1}^{r}|a_j\rangle\langle a_j|.$
Since $\operatorname{Ran}(Q)\subseteq \operatorname{Ran}(Q_A\otimes I_B)$, one has
$Q\le Q_A\otimes I_B.$
Hence
\begin{equation*}
\langle b|Q|b\rangle
\le
\langle b|(Q_A\otimes I_B)|b\rangle
=
\operatorname{Tr}(Q_A\rho_A).
\end{equation*}
By the Ky Fan maximum principle\cite{Fan1951},
$\operatorname{Tr}(Q_A\rho_A)\le \sum_{j=1}^{r}\mu_j.$
Therefore
\begin{equation*}
\Omega_{\mathcal{G}_{A|B},\Pi_r^{\mathrm{prod}}}(|b\rangle)
\le
\sum_{j=1}^{r}\mu_j.
\end{equation*}
Combining the above relations, we obtain
\[
\Omega_{\mathcal{G}_{A|B},\Pi_r^{\mathrm{prod}}}(|b\rangle)
=
\max_{\mathrm{SR}(|\phi\rangle)\le r}
|\langle \phi|b\rangle|^2
=
\sum_{j=1}^{r}\mu_j.
\]
Hence, by the definition of $E_{g,r}^{A|B}$,
Eq.~(\ref{eq:sec5_thm4_P}) follows from Proposition~1, and the rank-one case is immediate. \qed


\begin{thebibliography}{S2}
\bibitem{SEC}
E. C. G. Sudarshan, Equivalence of semiclassical and quantum mechanical descriptions of statistical light beams, Phys. Rev. Lett. 10 (1963) 277.

\bibitem{GRJ}
R. J. Glauber, Coherent and incoherent states of the radiation field, Phys. Rev. 131 (1963) 2766.

\bibitem{SMO}
M. O. Scully, M. S. Zubairy, Quantum Optics, Cambridge University Press, Cambridge, 1997.

\bibitem{TB}
T. Baumgratz, M. Cramer, M. B. Plenio, Quantifying coherence, Phys. Rev. Lett. 113 (2014) 140401.

\bibitem{YXD}
X. Yu, D. Zhang, G. Xu, D. Tong, Alternative framework for quantifying coherence, Phys. Rev. A 94 (2016) 060302.

\bibitem{RobustnessCoh}
C. Napoli, T. R. Bromley, M. Cianciaruso, M. Piani, N. Johnston, G. Adesso, Robustness of coherence: an operational and observable measure of quantum coherence, Phys. Rev. Lett. 116 (2016) 150502.

\bibitem{SAS}
A. Streltsov, U. Singh, H. S. Dhar, M. N. Bera, G. Adesso, Measuring quantum coherence with entanglement, Phys. Rev. Lett. 115 (2015) 020403.

\bibitem{QXG}
X. Qi, T. Gao, F. Yan, Measuring coherence with entanglement concurrence, J. Phys. A: Math. Theor. 50 (2017) 285301.

\bibitem{DSB}
S. Du, Z. Bai, X. Qi, Coherence measures and optimal conversion for coherent states, Quantum Inf. Comput. 15 (2015) 1307--1316.

\bibitem{SDZ}
S. Du, Z. Bai, Y. Guo, Conditions for coherence transformations under incoherent operations, Phys. Rev. A 91 (2015) 052120.

\bibitem{TRB}
T. R. Bromley, M. Cianciaruso, G. Adesso, Frozen quantum coherence, Phys. Rev. Lett. 114 (2015) 210401.

\bibitem{JWZ}
J. Wang, Z. Tian, J. Jing, H. Fan, Irreversible degradation of quantum coherence under relativistic motion, Phys. Rev. A 93 (2016) 062105.

\bibitem{YWF-Grover}
L. Ye, Z. Wu, S. M. Fei, Tsallis relative $\alpha$ entropy of coherence dynamics in Grover's search algorithm, Commun. Theor. Phys. 75 (2023) 085101.

\bibitem{YWF-Linear}
L. Ye, Z. Wu, S. M. Fei, Coherence dynamics in quantum algorithm for linear systems of equations, Phys. Scr. 98 (2023) 125104.

\bibitem{XDY}
X. Yu, D. Zhang, C. Liu, D. Tong, Measure-independent freezing of quantum coherence, Phys. Rev. A 93 (2016) 060303.

\bibitem{YPYJ}
Y. Peng, Y. Jiang, H. Fan, Maximally coherent states and coherence-preserving operations, Phys. Rev. A 93 (2016) 032326.

\bibitem{AMV}
A. Mani, V. Karimipour, Cohering and decohering power of quantum channels, Phys. Rev. A 92 (2015) 032331.

\bibitem{CRM}
C. Radhakrishnan, M. Parthasarathy, S. Jambulingam, T. Byrnes, Distribution of quantum coherence in multipartite systems, Phys. Rev. Lett. 116 (2016) 150504.

\bibitem{YWF-Simon}
L. Ye, Z. Wu, S. M. Fei, Coherence dynamics in Simon's quantum algorithm, Europhys. Lett. 144 (2023) 18001.

\bibitem{YWF-Shor}
L. Ye, Z. Wu, S. M. Fei, Coherence and entanglement dynamics in Shor's algorithm, Commun. Theor. Phys. 78 (2026) 015102.

\bibitem{WZQH}
Z. Wu, H. Huang, S.-M. Fei, X. Li-Jost, Geometry of skew information-based quantum coherence, Commun. Theor. Phys. 72 (2020) 105102.

\bibitem{WZZL}
Z. Wu, L. Zhang, S.-M. Fei, X. Li-Jost, Average skew information-based coherence and its typicality for random quantum states, J. Phys. A: Math. Theor. 54 (2021) 015302.

\bibitem{YWZ-Circuit}
L. Ye, Z. Wu, N. Zhou, Coherence and imaginarity as resources in quantum circuit complexity, Adv. Quantum Technol. 9 (2026) e01007.

\bibitem{WZZF}
Z. Wu, L. Zhang, S.-M. Fei, J. Wang, Skew information-based coherence generating power of quantum channels, Quantum Inf. Process. 21 (2022) 236.

\bibitem{ECM}
E. Chitambar, M.-H. Hsieh, Relating the resource theories of entanglement and quantum coherence, Phys. Rev. Lett. 117 (2016) 020402.

\bibitem{JMB}
J. Ma, B. Yadin, D. Girolami, V. Vedral, M. Gu, Converting coherence to quantum correlations, Phys. Rev. Lett. 116 (2016) 160407.

\bibitem{ZXY}
Z. Xi, Y. Li, H. Fan, Quantum coherence and correlations in quantum system, Sci. Rep. 5 (2015) 10922.

\bibitem{YYX}
Y. Yao, X. Xiao, L. Ge, C. Sun, Quantum coherence in multipartite systems, Phys. Rev. A 92 (2015) 022112.

\bibitem{JXP}
J. Xu, Quantifying coherence of Gaussian states, Phys. Rev. A 93 (2016) 032111.

\bibitem{YRZ}
Y. Zhang, L. Shao, Y. Li, H. Fan, Quantifying coherence in infinite-dimensional systems, Phys. Rev. A 93 (2016) 012334.

\bibitem{Zhao2024GeoCoherence}
M. J. Zhao, T. Xia, L. Zhang, B. Li, Geometric coherence of some mixed states, Eur. Phys. J. Plus 139 (2024) 522.

\bibitem{Horodecki2009}
R. Horodecki, P. Horodecki, M. Horodecki, K. Horodecki, Quantum entanglement, Rev. Mod. Phys. 81 (2009) 865.

\bibitem{Bruss2002}
D. Bru{\ss}, Characterizing entanglement, J. Math. Phys. 43 (2002) 4237.

\bibitem{Bell1964}
J. S. Bell, On the Einstein Podolsky Rosen paradox, Physics Physique Fizika 1 (1964) 195.

\bibitem{Bennett1992}
C. H. Bennett, S. J. Wiesner, Communication via one- and two-particle operators on Einstein-Podolsky-Rosen states, Phys. Rev. Lett. 69 (1992) 2881.

\bibitem{Ekert1991}
A. K. Ekert, Quantum cryptography based on Bell's theorem, Phys. Rev. Lett. 67 (1991) 661.

\bibitem{Shimony1995}
A. Shimony, Degree of entanglement, Ann. N. Y. Acad. Sci. 755 (1995) 675--679.

\bibitem{Wei2003}
T. C. Wei, P. M. Goldbart, Geometric measure of entanglement and applications to bipartite and multipartite quantum states, Phys. Rev. A 68 (2003) 042307.

\bibitem{Guhne2007}
O. G\"uhne, M. Reimpell, R. F. Werner, Estimating entanglement measures in experiments, Phys. Rev. Lett. 98 (2007) 110502.

\bibitem{Bravyi2025QECGeometric}
S. Bravyi, D. Lee, Z. Li, B. Yoshida, How much entanglement is needed for quantum error correction?, Phys. Rev. Lett. 134 (2025) 210602.

\bibitem{Namiki1991}
M. Namiki, S. Pascazio, Wave-function collapse by measurement and its simulation, Phys. Rev. A 44 (1991) 39.

\bibitem{Ricardo2020}
R. Karam, Why are complex numbers needed in quantum mechanics? Some answers for the introductory level, Am. J. Phys. 88 (2020) 39--45.

\bibitem{David2018}
D. Gaspard, Connection formulas between Coulomb wave functions, J. Math. Phys. 59 (2018) 112104.

\bibitem{Hickey2018}
A. Hickey, G. Gour, Quantifying the imaginarity of quantum mechanics, J. Phys. A: Math. Theor. 51 (2018) 414009.

\bibitem{Wu2021}
K. D. Wu, T. V. Kondra, S. Rana, C. M. Scandolo, G. Y. Xiang, C. F. Li, G. C. Guo, A. Streltsov, Resource theory of imaginarity: quantification and state conversion, Phys. Rev. A 103 (2021) 032401.

\bibitem{Guo2025GeoLikeImaginarity}
M. L. Guo, B. Li, S. M. Fei, Geometric-like imaginarity: quantification and state conversion, Sci. China Phys. Mech. Astron. 68 (2025) 220311.

\bibitem{WuWu2025CTP}
C. Wu, Z. Wu, Two imaginarity monotones induced by unified ($\alpha$, $\beta$)-relative entropy, Commun. Theor. Phys. 77 (2025) 095101.

\bibitem{GMG-AQT}
M. L. Guo, B. Li, B. Hu, S. M. Fei,
Quantifying the imaginarity via different distance measures,
Adv. Quantum Technol. 8 (2025) 2400562.

\bibitem{GMG-PRA}
M. L. Guo, N. Guan, B. Li, B. Hu, S. M. Fei,
Imaginarity as a resource within quantum coherence: Geometric decomposition and operational conversion,
Phys. Rev. A 114 (2026) 012435.

\bibitem{MLi2026}
M.-S. Li, R. Wagner, L. Zhang,
Multistate imaginarity and coherence in qubit systems, Phys. Rev. A 113 (2026) 012428.

\bibitem{Guo2026SetImaginarity}
Y. Guo, J. Pan, Y. Wang, S. Du, Measure of set imaginarity, Phys. Rev. A 114 (2026) 022453.

\bibitem{WW-Operations}
C. Wu, Z. Wu, Quantifying imaginarity of quantum operations, Sci. China Phys. Mech. Astron. 69 (2026) 230316.

\bibitem{WMH-PRA}
S. M. Wang, M. L. Hu, H. Fan,
Off-resonant preservation and generation of imaginarity in distributed scenarios,
Phys. Rev. A 113 (2026) 052457.

\bibitem{HAW}
A. W. Harrow, A. Hassidim, S. Lloyd, Quantum algorithm for linear systems of equations, Phys. Rev. Lett. 103 (2009) 150502.

\bibitem{Kerenidis}
I. Kerenidis, A. Prakash, Quantum recommendation systems, in: 8th Innovations in Theoretical Computer Science Conference (ITCS 2017), Leibniz Int. Proc. Inform. 67 (2017) 49:1--49:21.

\bibitem{Wossnig}
L. Wossnig, Z. Zhao, A. Prakash, Quantum linear system algorithm for dense matrices, Phys. Rev. Lett. 120 (2018) 050502.

\bibitem{Papagiannis2025QSolvers}
N. Papagiannis, M. Vavalis, On quantum solvers for linear algebraic systems, Quantum Inf. Comput. 25 (2025) 640--667.

%\bibitem{Childs2017}
%A. M. Childs, R. Kothari, R. D. Somma, Quantum algorithm for systems of linear equations with exponentially improved dependence on precision, SIAM J. Comput. 46 (2017) 1920.

\bibitem{ZhangLiYuan2022QSP}
X. M. Zhang, T. Li, X. Yuan, Quantum state preparation with optimal circuit depth: implementations and applications, Phys. Rev. Lett. 129 (2022) 230504.

\bibitem{ZhangYuan2026FS}
Y. Zhang, X. Yuan, Quantum algorithms for fidelity susceptibility: from quantum criticality to metrology, Phys. Rev. Lett. 136 (2026) 080604.

\bibitem{King2026SOSSA}
R. King, G. H. Low, R. Babbush, R. D. Somma, N. C. Rubin, Quantum simulation with sum-of-squares spectral amplification, Phys. Rev. Lett. 136 (2026) 110601.

\bibitem{Zhang2025NonHermitian}
X. M. Zhang, Y. Zhang, W. He, X. Yuan, Exponential quantum advantages for practical non-Hermitian eigenproblems, Phys. Rev. Lett. 135 (2025) 140601.

\bibitem{Zhang2026NonlinearProperties}
Y. Zhang, Y. Wu, Y. Zhou, X. Yuan, Measuring less to learn more: quadratic speedup in learning nonlinear properties of quantum states, Phys. Rev. Lett. 136 (2026) 130602.

\bibitem{Nguyen2022DenseKernels}
Q. T. Nguyen, B. T. Kiani, S. Lloyd, Block-encoding dense and full-rank kernels using hierarchical matrices: applications in quantum numerical linear algebra, Quantum 6 (2022) 876.

\bibitem{BravoPrieto2023VQLS}
C. Bravo-Prieto, R. LaRose, M. Cerezo, Y. Subasi, L. Cincio, P. J. Coles, Variational quantum linear solver, Quantum 7 (2023) 1188.

\bibitem{Ghisoni2026SQLS}
F. Ghisoni, F. Scala, D. Bajoni, D. Gerace, Resource-efficient quantum algorithm for linear systems of equations, Phys. Rev. A 113 (2026) 032405.

\bibitem{Gong2024QPSO}
C. Gong, N. R. Zhou, S. Xia, S. Huang, Quantum particle swarm optimization algorithm based on diversity migration strategy, Future Gener. Comput. Syst. 157 (2024) 445--458.

\bibitem{Gong2024QKNNDC}
L. H. Gong, W. Ding, Z. Li, Y. Z. Wang, N. R. Zhou, Quantum k-nearest neighbor classification algorithm via a divide-and-conquer strategy, Adv. Quantum Technol. 7 (2024) 2300221.

\bibitem{Gong2024QCNNVQC}
L. H. Gong, J. J. Pei, T. F. Zhang, N. R. Zhou, Quantum convolutional neural network based on variational quantum circuits, Opt. Commun. 550 (2024) 129993.

\bibitem{Huang2023HQCNN}
S. Y. Huang, W. J. An, D. S. Zhang, N. R. Zhou, Image classification and adversarial robustness analysis based on hybrid quantum-classical convolutional neural network, Opt. Commun. 533 (2023) 129287.

\bibitem{MA}
M. A. Nielsen, I. L. Chuang, Quantum Computation and Quantum Information, Cambridge University Press, Cambridge, 2000.

\bibitem{Wang2020}
C. Wang, L. Wossnig, A quantum algorithm for simulating non-sparse Hamiltonians, Quantum Inf. Comput. 20 (2020) 597--615.

\bibitem{LiWLF2019}
G. Li, Y. Wang, Y. Luo, Y. Feng, Quantum data fitting algorithm for non-sparse matrices, arXiv preprint arXiv:1907.06949, 2019.


\bibitem{Shor1997}
P. W. Shor, Polynomial-time algorithms for prime factorization and discrete logarithms on a quantum computer, SIAM J. Comput. 26 (1997) 1484--1509.

\bibitem{Bartlett2007}
S. D. Bartlett, T. Rudolph, R. W. Spekkens, Reference frames, superselection rules, and quantum information, Rev. Mod. Phys. 79 (2007) 555--609.

\bibitem{Zhou2024CohFraction}
S.-Q. Zhou, H. Jin, J.-M. Liang, S.-M. Fei, Y. Xiao, Z. Ma, Coherence fraction in Grover's search algorithm, Phys. Rev. A 110 (2024) 062429.

\bibitem{Chitambar2016}
E. Chitambar, G. Gour, Critical examination of incoherent operations and a physically consistent resource theory of quantum coherence, Phys. Rev. Lett. 117 (2016) 030401.

\bibitem{Fan1951}
K. Fan, Maximum properties and inequalities for the eigenvalues of completely continuous operators, Proc. Natl. Acad. Sci. U.S.A. 37 (1951) 760--766.


\end{thebibliography}
\end{document}